\documentclass[pre,aps,floats, superscriptaddress,longbibliography, twocolumn]{revtex4-1}
\usepackage{amsmath}
\usepackage{amssymb}
\usepackage{graphics}
\usepackage{graphicx}
\usepackage{hyperref}
\usepackage{bm}
\usepackage{wrapfig}

\usepackage{amsmath}
\usepackage{tikz}
\usetikzlibrary{shapes,decorations,calc}

\definecolor{dred}{rgb}{0.6,0.,0.}
\definecolor{dblue}{rgb}{0.,0.,0.5}

\hypersetup{colorlinks,linkcolor=blue,urlcolor=blue,citecolor=blue}

\begin{document} 

\title{Entanglement spectroscopy of chiral edge modes in the Quantum Hall effect}
\date{\today}
\author{Benoit Estienne}
\affiliation{Sorbonne Universit\'{e}, CNRS, Laboratoire de   Physique Th\'{e}orique et Hautes \'{E}nergies, LPTHE, F-75005 Paris, France.}
\author{Jean-Marie St\'ephan}
\affiliation{Univ Lyon, CNRS, Universit\'e Claude Bernard Lyon 1, UMR5208, Institut Camille Jordan, F-69622 Villeurbanne, France}

\begin{abstract}
	We investigate the entanglement entropy in the Integer Quantum Hall effect in the presence of an edge, performing an exact calculation directly from the microscopic two-dimensional wavefunction. The edge contribution is shown to coincide exactly with that of a chiral Dirac fermion, and this correspondence holds for an arbitrary collection of intervals. In particular for a single interval the celebrated conformal formula is recovered with left and right central charges $c+\bar{c} =1$.  Using Monte-Carlo techniques we establish  that this behavior persists for strongly interacting systems such as Laughlin liquids. This illustrates how entanglement entropy is  not only capable of detecting the presence of massless degrees of freedom, but also of pinpointing their position in real space, as well as elucidating their nature. 
\end{abstract}

\maketitle

\section{Introduction}

Quantum entanglement has proven to be a fundamental and versatile tool to reveal the physical properties of condensed matter systems. In particular, the scaling of the entanglement entropy (EE) can unveil most of the long-distance properties of a system. For example, it provides a direct method to extract the central charge in one-dimensional critical systems through the celebrated log law\cite{Holzhey,Vidal_ee1d,CalabreseCardy_2004}. For a gapped system in any dimension, the EE obeys the area law:  its leading term grows like the size of the boundary between two subsystems instead of their volume. In two-dimensions, corrections to the area law can be used to detect \emph{tolopological order} and even measure the quantum dimension of the various anyonic excitations \cite{KitaevPreskill,LevinWen}. Moreover in topological matter entanglement spectroscopy is tied to the physical conducting surface states in a mechanism known as \emph{bulk-edge correspondence} \cite{PhysRevLett.101.010504,PhysRevB.84.205136,Patlatiuk}.

The observation of the quantum Hall effect (QHE) has been a major discovery in the 1980s, and marked the discovery of topological quantum matter. It is perhaps the simplest setup in which topological order can emerge, and thus serves as a paradigmatic system. The scaling of the EE in the insulating 'bulk' of a quantum Hall liquid is well understood \cite{RodriguezSierra,RodriguezSierra2}, and the area law in this context is even a mathematical theorem \cite{arealaw_proof}. A distinctive feature of topological phases is that while they are insulators in the bulk, they display exotic, topologically protected metallic states at their surfaces/edges. The bulk being gapped, these edge states are responsible for the non-trivial transport properties at low temperatures. In the particular case of the QHE, these modes are \emph{chiral} : they can only propagate in one direction. 

In this letter we investigate how the (unreconstructed) edge modes affects the EE in a quantum Hall liquid. It is well known that the low-energy effective description of these edge modes is a one-dimensional chiral Dirac fermion. However their contribution to the EE is a rather non-trivial question. In particular these conducting modes exhibit a \emph{chiral anomaly}. This anomaly together with charge conservation implies that such edge modes cannot appear in a strictly one-dimensional system : they can only appear at the edge of a two-dimensional one - in our case, a quantum Hall liquid. Furthermore there are non-trivial correlations between the bulk and the edge degrees of freedom. 

We perform an exact calculation directly from the microscopic two-dimensional wavefunction of the integer quantum Hall liquid. We show that the edge contribution to the EE is given exactly by that of a one-dimensional chiral Dirac fermion by providing explicit results for an arbitrary collection of intervals.  In particular if the intersection of the region $A$ with the edge is a single interval of size $l$, we recover the celebrated logarithmic behavior \cite{Holzhey,Vidal_ee1d,CalabreseCardy_2004}
\begin{align}
S_{\textrm{edge}} \simeq \frac{c+\bar{c}}{6} \log l, \qquad c+\bar{c} =1
\end{align}
Using entanglement entropy to numerically extract the central charge of gapless modes has already been used in several numerical studies \cite{PhysRevB.88.155314,Crepel1,Crepel2,Crepel3}. While these studies were reasonably conclusive, they involved heavy matrix product states machinery, leading to important finite size effects. In order to demonstrate the validity of this approach, we provide here a case study in a much simpler setup  - namely the interface between the integer quantum hall liquid with the vacuum - which allows for a complete analytic derivation. We also establish a refined result in the multi-interval case. To further benchmark this method we finally consider an interface between a Laughlin state and the vacuum, which can be analysed for very large systems using Monte Carlo methods. 

\section{Setup and Methodology}

We are concerned with a two-dimensional electron gas subject to a perpendicular magnetic field $B$. We implicitly assume the presence of a confining potential $V(x,y)$ whose only role is to lift the lowest Landau level (LLL) degeneracy, thus ensuring that the many-body ground-state has a well-defined edge. This potential is assumed to be smooth and very weak compared to the cyclotron gap, thus neglecting Landau level mixing. This perturbative treatment is merely here to simplify the discussion and is not crucial to our results. For instance it is rather straightforward to solve exactly the case of a quadratic confining potential.  The only effect is a slight modification of the wavefunctions in the lowest band, and our calculation goes through essentially unhindered. We further assume that electron-electron interactions are sufficiently weak to avoid edge reconstruction.

For large magnetic field the typical distance over which the potential varies is much larger than the magnetic length scale $l_B = \sqrt{\hbar/eB}$. In practice this means that we work in the semi-classical regime $l_B \to 0$. This is the relevant regime to study the universal behavior of the entanglement entropy, that is for regions much larger than all microscopic scales. 

Our analytic derivation rests on two facts. First, the integer quantum Hall effect can be understood using free electrons, for which there are extremely efficient methods to compute the EE \cite{Peschel_2003}. In particular the R\'enyi entropies are related to charge fluctuations in the subsystem $A$ through
\begin{align}
S_n = \sum_m s_n(m) \kappa_{2m}
\end{align}
where $s_n(m)$ are numerical coefficients independent of the problem studied and $\kappa_{2m}$ are the even cumulants of the particle-number distribution in region $A$ \cite{Klich_2006,Klich_2009,Calabrese_2012}. In particular $\kappa = \kappa_2$ is the variance of the number of particles in region $A$.  Second, in cases where the higher cumulants are suppressed relatively to the particle variance, the above series can be truncated to its first term, and the asymptotic behavior of the R\'enyi entropies is proportional to the variance $\kappa$ of the particle number
\begin{align}\label{eq:relationC2S}
S_n =  \frac{\pi^2}{6}\left(1+ \frac{1}{n} \right) \kappa  + \textrm{subleading terms} \,.
\end{align}
While in the bulk of a QH droplet this condition is not satisfied, the charge fluctuations at the edge of a QH droplet are expected to be Gaussian. In the case of a single interval in the strip geometry (as in fig. \eqref{Fig:Confining_potential0}) we are also able to compute all the cumulants asymptotically, and prove the gaussianity of charge fluctuations. For other geometries we will simply assume that \eqref{eq:relationC2S} holds.\\

The problem of extracting the edge contribution to the EE in the QHE thus boil down to the computation of the  variance $\kappa$. If we denote by $\{\phi_i, i \in I\}$ the occupied one-body states, then the particle variance in region $A$ is simply given by \cite{Calabrese_2012,arealaw_proof} 
\begin{align}
\kappa = \int_{A} d^2r_1\int_{A^c} d^2r_2 \left| K(r_1,r_2)  \right|^2 \,,
\end{align}
where $K(r_1,r_2)=\sum_{i \in I} \phi_i(r_1) \phi^*_i(r_2)$ is the kernel of the projector to the occupied states, $r_j=(x_j,y_j)$ is the position in the plane. In the following we consider two simple geometries in which the above kernel can be computed exactly, namely the half-plane and the disk. 
It is then a simpler matter to extract the asymptotic behavior in the semi-classical regime $l_B \to 0)$ of the variance.

\section{Infinite strip geometry}

\begin{figure}[h]
	\centering
	\includegraphics[width=0.7\columnwidth]{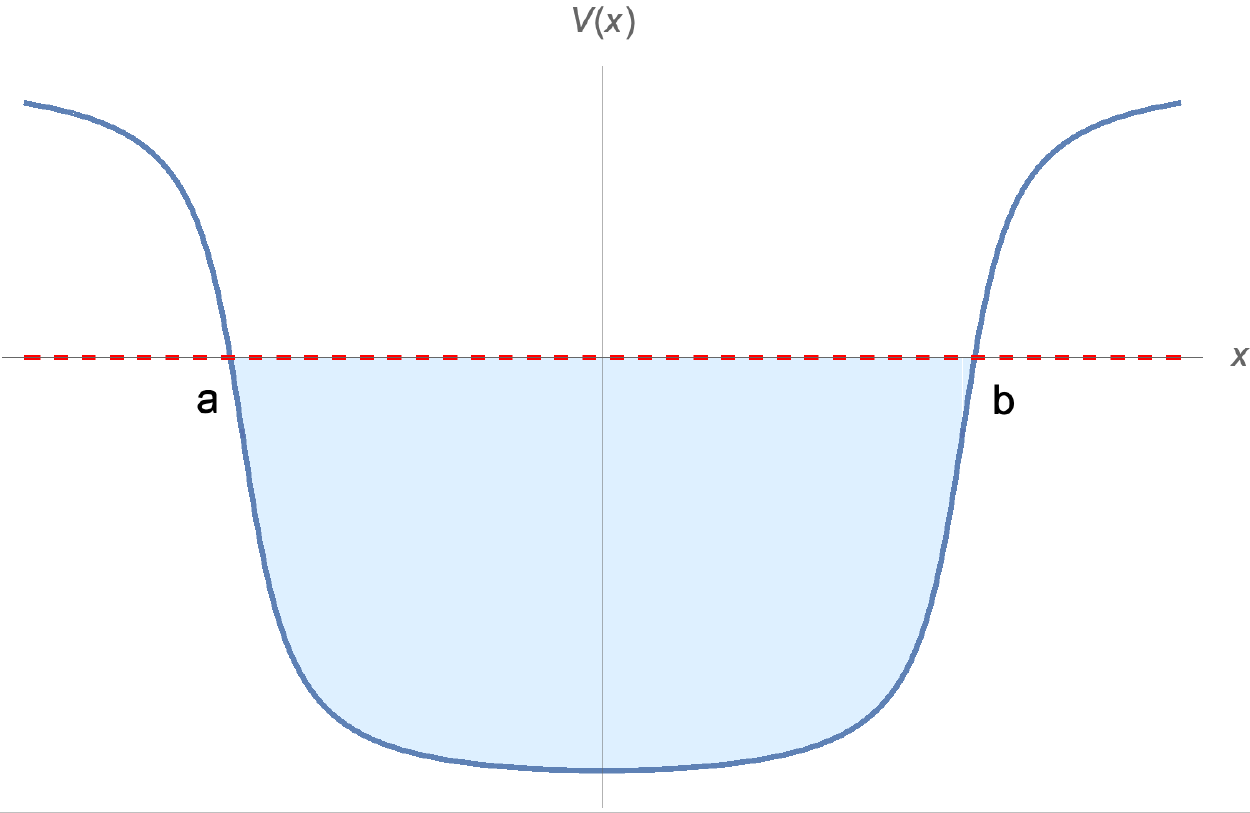}\includegraphics[width=0.3\columnwidth]{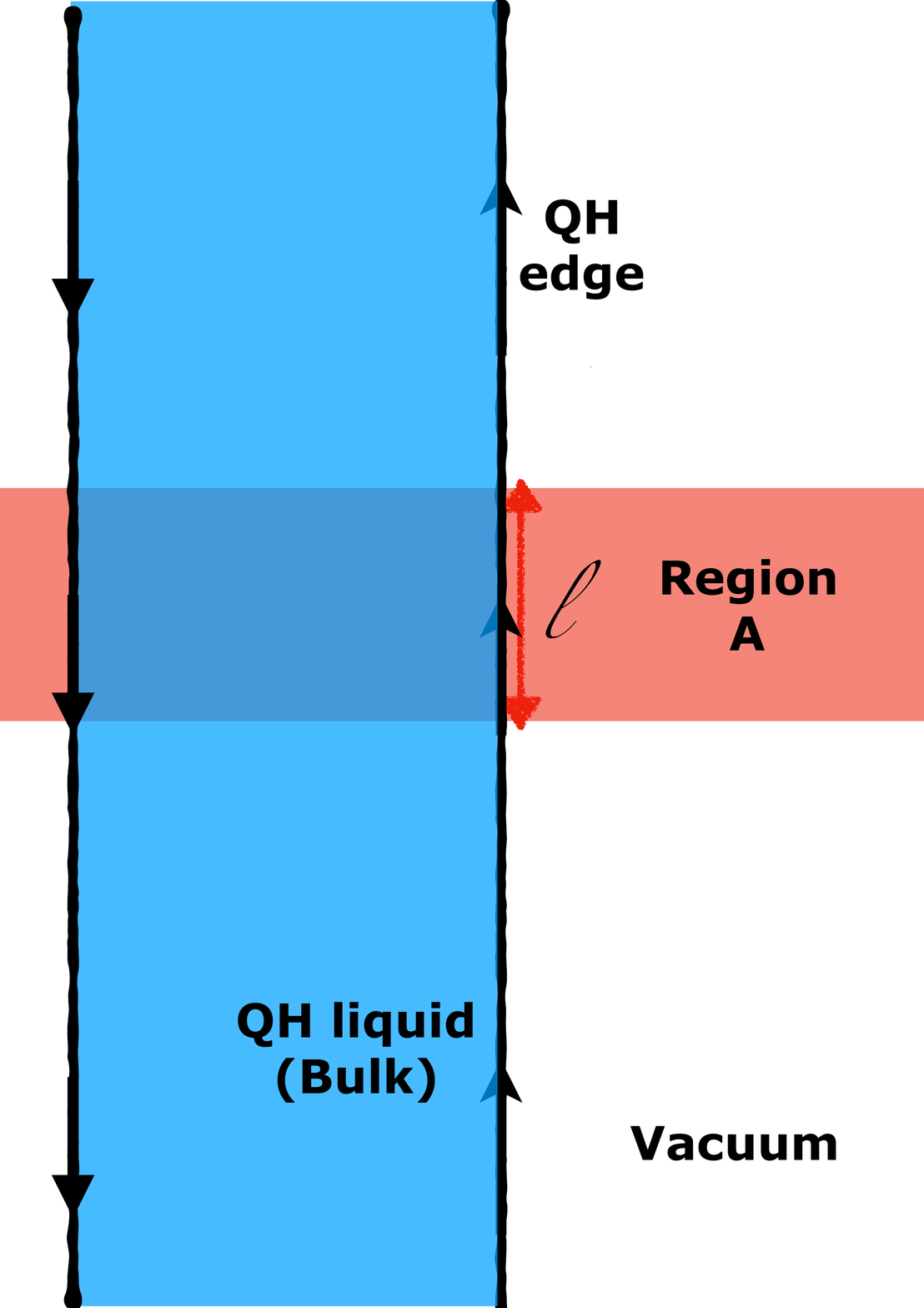}
  \caption{Left : sketch of the potential $V(x)$ used to confine the Quantum Hall liquid to the strip $a < x <b$. Right : in blue, the quantum Hall droplet in the strip geometry with its left and right chiral edge modes. In red, the region $A$ used for the bipartition defining the entanglement entropy.}
	\label{Fig:Confining_potential0}
\end{figure}

We work on the plane, in the Landau gauge $A = B \left( \begin{array}{c} 0 \\ x \end{array} \right)$
in which case a "basis" of the LLL wavefunctions is given by 
\begin{align}
\phi_q(x,y) =  \frac{1}{\sqrt{l_B \sqrt{\pi} }}e^{i \frac{q y}{l_B^2}} e^{- \frac{(x - q)^2}{2 l_B^2}}
\end{align}
with normalization $\langle \phi_q | \phi_{q'} \rangle = 2\pi l_B^2  \delta(q-q')$.
We create edges by imposing a smooth confining potential $V = V(x)$.
With a potential as in fig.~\eqref{Fig:Confining_potential0} one creates a quantum Hall droplet occupying (in the semi-classical limit $l_B \to 0$) the semi-infinite strip $a < x < b$. 
 The corresponding kernel is $K(r_1,r_2) =  \int_{a}^{b} \frac{dq}{2\pi l_B^2} \,  \phi_q(r_1)\phi^*_q(r_2)$, \emph{i.e.}
\begin{align}
K(r_1,r_2)  =  \frac{1}{2\pi\sqrt{\pi} l^3_B} e^{- \frac{x_1^2 + x_2^2}{2l_B^2}} \int_{a}^{b} dq \, e^{\frac{q (z_1+\bar{z}_2)}{l_B^2}}   e^{-\frac{q^2}{l_B^2}} \label{strip_kernel} 
\end{align}
where $z_j = x_j + i y_j$. 
This integral can be computed in terms of the error function, however the integral form is more useful than the explicit expression for the purpose of extracting its asymptotic behavior. 
Note that we recover the usual Bergman kernel on the plane (in the Landau gauge) for $a = -\infty$, $b=\infty$. 

We have specified the geometry of the quantum Hall droplet, we now need to choose a subregion $A$ that intersects the edges. The simplest choice is a region $A$ which is invariant under $x$ translations, \emph{i.e.} 
\begin{align*}
A = \mathbb{R} \times \mathcal{A}
\end{align*}
as in  fig.~\eqref{Fig:Confining_potential0}. 
We will restrict ourselves to the case when $\mathcal{A}$ is a finite reunion of disjoint, finite length, intervals, say $\mathcal{A} = I_1 \cup \cdots \cup I_p$.  In that case 
\begin{align}
 \kappa  =  \int_{\mathcal{A}\times\mathcal{A}^c}  \frac{dy_1 dy_2}{4 \pi^2 l_B^4}  \,      \int_{[a,b]^2} dq_1 dq_2    \,   e^{i \frac{ (q_1-q_2) (y_1-y_2)}{l_B^2}}  e^{-\frac{(q_1-q_2)^2}{2l_B^2}}   
\end{align}

\begin{figure}[h]
	\centering
	\includegraphics[width=0.35\columnwidth]{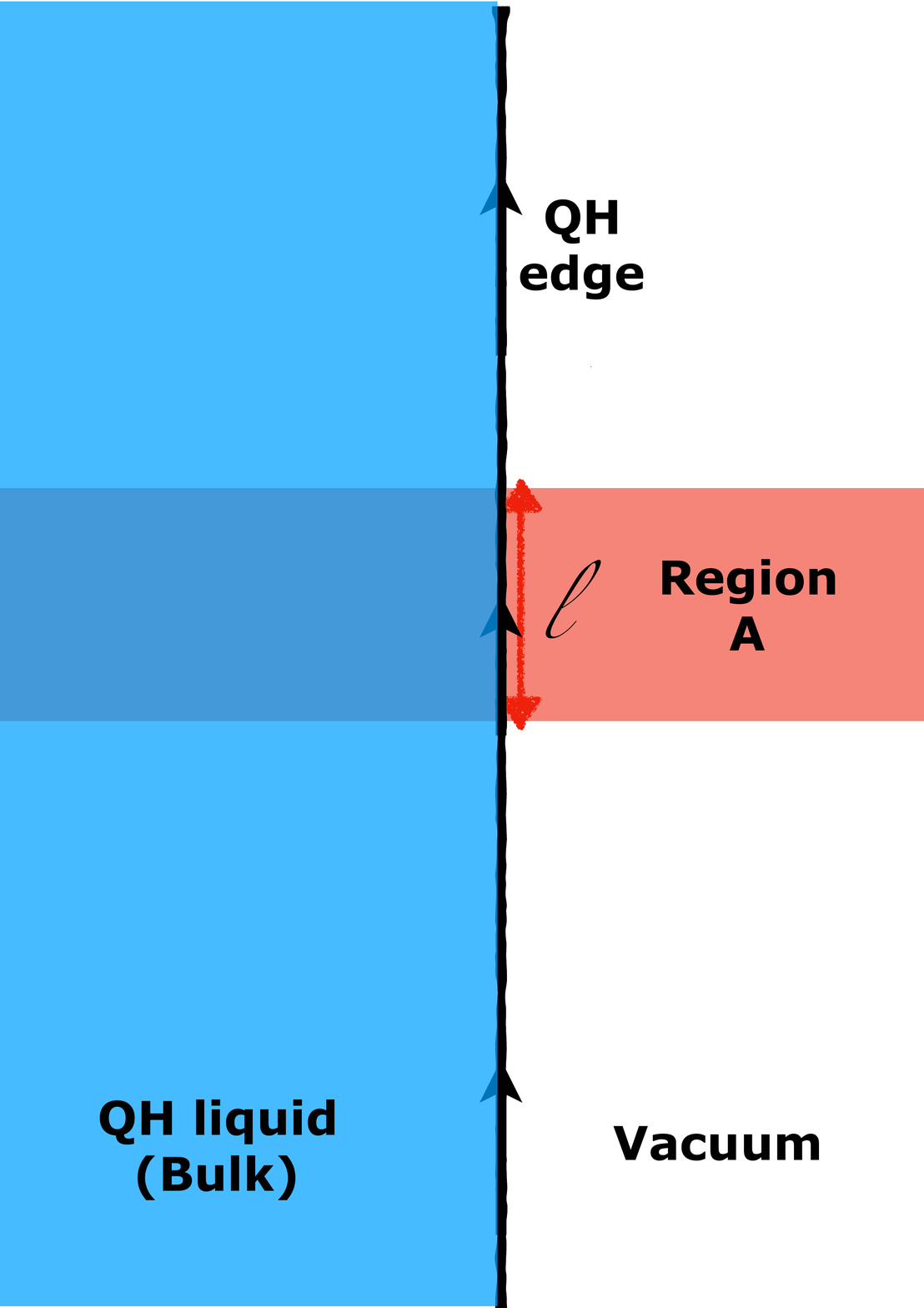}\hspace{1.5cm}\includegraphics[width=0.35\columnwidth]{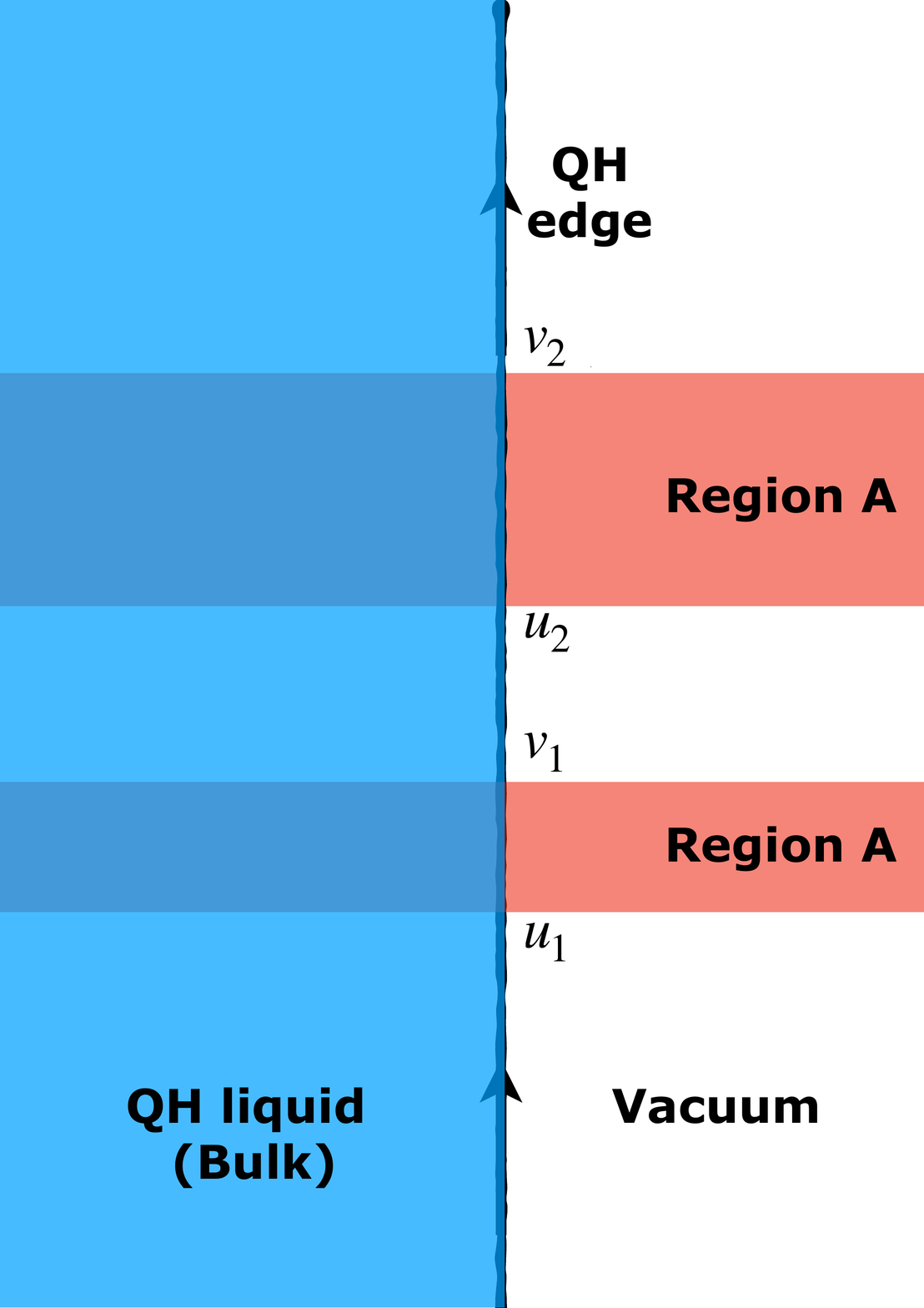}
		  \caption{In blue, the quantum Hall droplet in the half-plane geometry with its right chiral edge mode. In red, the region $A$ used for the bipartition defining the entanglement entropy. Left : single interval case. Right : two intervals.}
	\label{Fig:Confining_potential}
\end{figure}

The geometry of the strip has the drawback of having two edges, one at $x=a$ and the other at $x=b$. In order to extract the contribution of a single chiral edge mode, we consider a QH droplet occupying the half-plane $\Gamma = \{ (x,y) \in \mathbb{R}^2, \, x <0 \}$ as in fig.~\eqref{Fig:Confining_potential}. 
Formally this means sending $a$ to $-\infty$, however taking this limit would yield a diverging entropy/variance caused by the bulk area law. 
A simple way to remove this diverging piece is to consider 
\begin{align}
\Delta \kappa =   \frac{\kappa (\Gamma) + \kappa(\Gamma^c) -  \kappa(\mathbb{R}^2)}{2}   
\end{align}
By inversion symmetry $\kappa (\Gamma) =  \kappa (\Gamma^c)$. Moreover $ \kappa(\mathbb{R}^2)$ gives a pure area law, with no algebraic correction. So this subtraction scheme singles out the edge contribution by removing the area law term.  After a few elementary manipulations one ends up with
a formula that is extremely suggestive of one-dimensional free fermions : 
\begin{align}
\Delta \kappa& =   \int_{\mathcal{A} \times\mathcal{A}^c} dy_1 dy_2 \,  \mathcal{K}_{l_B}(y_1-y_2), 
\end{align}
where $\mathcal{K}_{l_B}(y)  =  \frac{1}{l_B^2} \mathcal{K}(y/l_B)$  is given by
\begin{align}
\mathcal{K}(y) & = - \frac{1}{4 \pi^2}  \int_{0}^{\infty}  \,   t e^{-\frac{t^2}{2}}   \cos ty\, dt   \\
 & = \frac{1}{4\pi^2} \left( \sqrt{2} y F\left(\frac{y}{\sqrt{2}}\right) -1 \right)
 \label{strip_effective_kernel}
 \end{align}
where  $F(x) = e^{-x^2} \int_{0}^x e^{t^2} dt$  is the Dawson function. One can think of $\mathcal{K}_{l_B}(y)$ as the square of the effective propagator of the edge chiral fermion, in which the magnetic length $l_B$ plays the role of a short-distance cutoff. At long distances ($y \gg l_B$) it decays algebraically as  $\mathcal{K}_{l_B}(y) \sim (2\pi y)^{-2}$, which is the expected exponent for a critical fermion in 1+1 dimensions.

When $\mathcal{A}$ is a single interval, whose length we denote by $l$, as in  fig.~\eqref{Fig:Confining_potential}, we find 
 \begin{align}
 \label{strip_plane_exact}
\Delta \kappa  & = \frac{1}{2\pi^2} \int_{0}^{\infty} \, e^{-\frac{t^2}{2}} \frac{1 - \cos \frac{l}{l_B} t}{t} dt  \\
 &  =   \frac{l^2}{4\pi^2 l_B^2} \, _2F_2\left(1,1;\frac{3}{2},2;-\frac{l^2}{2 l_B^2}\right).
\end{align}
In the relevant regime $l \gg l_B$ this yields
 \begin{align}
\Delta \kappa  =   \frac{1}{2\pi^2} \log \frac{l}{l_B}  + \frac{\log 2+\gamma}{4 \pi ^2}-\frac{l_B^2}{4 \pi ^2 l^2} +O\left(\left(\frac{l_B}{l}\right)^4\right)
\end{align}
where $\gamma$ is Euler's constant. We prove in appendix \ref{app:C0} that all higher order cumulants are suppressed with respect to variance. More previsely, we find
\begin{align}
 \Delta \kappa_{2n}=O(1)\qquad,\qquad \Delta \kappa_{2n-1}=O(l_B/l),
\end{align}
for $n\geq 2$. This implies that the edge entanglement entropy behaves as
\begin{align}
\Delta S_n \sim \frac{1}{12}\left(1+\frac{1}{n}\right)\log \frac{l}{l_B},
\end{align}
which is compatible with the celebrated conformal result
\begin{align}
S \sim \frac{c+\bar{c}}{12}\left(1+\frac{1}{n}\right)  \log l 
\end{align}
for a chiral Dirac fermion, for which $c=1$ and $\bar{c} =0$. But from the single interval EE one can only conclude that $c + \bar{c} =1$. To unequivocally identify the universality class of the edge, one possibility is to repeat the above calculation  in the case of several intervals as in fig.~\eqref{Fig:Confining_potential}. When $\mathcal{A} = \cup_{a=1}^p I_a$, where $I_a = [u_a,v_a]$ are  (compact) disjoint intervals, a simple calculation (see Appendix \ref{app:C1}) yields
\begin{align}
\label{eq_multiple_intervals_half_plane}
\Delta S_n \sim \frac{1}{12}\left(1+\frac{1}{n}\right) \log \prod_{a=1}^p \frac{v_a -u_a}{l_B} \prod_{a<b} \frac{(u_b - v_a)(v_b - u_a)}{(u_b-u_a)(v_b-v_a)}
\end{align}
and we recover (half) the usual expression for the entanglement entropy of a complex fermion \cite{Casini_2005,Casini_2009}.

More generally the low-energy effective quantum field theory of the edge excitations of a fractional quantum Hall liquid is supposed to be a conformal field theory \cite{PhysRevB.79.245304,Zabrodin_Wiegmann}. It is not immediately clear what would be the analogue of \eqref{eq_multiple_intervals_half_plane} for states such as the Laughlin or the Moore-Read state. Indeed for a critical system in 1+1 dimensions, the R\'enyi entropies $S_n$ for $p$ intervals is given by the partition function on a Riemann surface of genus $(n-1)(p-1)$, or equivalently by a $2p$-point correlation function of twist fields in a cyclic orbifold \cite{Caraglio_2008,PhysRevLett.102.170602,Calabrese_2009,Calabrese_2011}. Generically such an object only makes sense for a full conformal field theory with $c = \bar{c}$. At the chiral level, one has instead a vector space of conformal blocks \cite{Dupic}. A natural conjecture is that the edge contribution to the entanglement entropy is a particular linear combination of these conformal blocks, very much like correlation functions in a boundary conformal field theory. In particular one would expect the edge entanglement entropy depends non-trivially on the bulk topological sector. 

To conclude this section, let us finally mention that the result can easily be generalized to regions which intersect the quantum Hall droplet with an angle different from $\pi/2$. This is discussed in appendix~\ref{app:corner}.

\section{The disk geometry}

We also study the original Laughlin wave function for $N$ particles, which is given in the symmetric gauge by
\begin{equation}\label{eq:Laughlin}
 \psi(z_1,\ldots,z_N)=\frac{1}{\sqrt{Z_N(\alpha)}}\prod_{1\leq i<j\leq N}(z_i-z_j)^\alpha e^{-\frac{1}{2}\sum_{i=1}^N |z_i|^2}
\end{equation}
where $\alpha$ is, in general, any positive integer. The constant factor $Z_N(\alpha)$ ensures normalization. Here the magnetic length has been set to $l_B=1/\sqrt{2}$, so that the semiclassical limit is achieved by taking $N\to \infty$. From standard arguments, the density is essentially constant inside a disk of radius $R=\sqrt{\alpha N}$ for large $N$, and zero outside. The edge is described by a chiral $U(1)$ CFT with central charge $c=1$ and Luttinger parameter $K=1/\alpha$. We note that this geometry is slightly more complicated than the previous one. In particular, we were not able to find a proof of gaussianity, even though we believe it is possible. 
\subsection{Analytical results for free fermions case $\alpha=1$}
\label{sec:disk_calculations}
As is well known the particular value $\alpha=1$ corresponds to free fermions, so we can use the techniques described in the previous section to study entropy and fluctuations (other values of $\alpha$ will be studied numerically in the next subsection). For the subsystem $A$ we choose the angular sector $A=\{z\in \mathbb{C},0<\arg z<\theta \}$. Contrary to Ref.~\cite{Petrescu_2014,Toine} this geometry breaks rotational symmetry, which is important for our purpose. It also has the feature of intersecting the edge of the droplet with an angle ($\pi/2$) which does not depend on $\theta$, sidestepping the issues discussed in appendix \ref{app:corner}. As before everything is encoded in the correlation kernel, which reads
\begin{equation}
  V_N(z,w)=\frac{1}{\pi}e^{-\frac{|z|^2+|w|^2-2z^* w}{2}}\frac{\Gamma(N,z^* w)}{\Gamma(N)}
\end{equation}
where $\Gamma(N,\alpha)$ is the incomplete Gamma function
\begin{eqnarray}
 \Gamma(N,\alpha)=\int_\alpha^\infty t^{N-1}e^{-t}dt,
\end{eqnarray}
and $\Gamma(N)=\Gamma(N,0)$ the usual Gamma function. 
It is possible to evaluate exactly the second cumulant asymptotically using this, see the calculation presented in appendix~\ref{app:C2}. Let us however present a short version which illustrates the physical origin of the angle dependence, which is expected from CFT. Instead of computing the second cumulant directly, we focus on its second derivative with respect to $\theta$. This has the advantage of getting rid of the area law term, which is linear in $\theta$ in this geometry. We find
\begin{equation}\label{eq:c2d2}
 \frac{d^2 \kappa(\theta)}{d\theta^2}=-2\int_0^\infty r\,dr \int_0^\infty s\,ds |V_N(r,se^{i\theta})|^2.
\end{equation}
For large $N$, the integrant in the previous equation is very small, except in two regions: (i) when $r$ and $s$ are small, in which case $V$ is given by the Bergman  kernel. We call this the corner contribution (ii) when $r,s$ are of order $\sqrt{N}$, in which case $V$ behaves like the chiral Dirac propagator. This is the CFT contribution. Since the two regions are well separated, we can write 
\begin{equation}
 \frac{d^2\kappa}{d\theta^2}=C(\theta)+E(\theta),
\end{equation}
where $C(\theta)$ and $E(\theta)$ are the corner and edge contributions, and compute them separately. For the first we use
\begin{equation}
 |V_N(r,s e^{i\theta})|^2=\frac{1}{\pi^2}e^{-r^2-s^2+2rs \cos \theta}
\end{equation}
for $\sqrt{N}-r\gg 1$ (same for $s$). A direct computation of the resulting integrals yields
\begin{equation}
C(\theta)=-\frac{1+(\pi-\theta)\cot \theta}{2\pi^2\sin^2\theta}.
\end{equation}
For the second one we use the asymptotic edge result,
\begin{equation}
 V(\sqrt{N}+\rho,(\sqrt{N}+\sigma)e^{i\theta})\underset{N\to \infty}{\sim} V_{\rm edge}(\rho,\sigma,\theta),
\end{equation}
where
\begin{equation}\label{eq:edgekernel}
 V_{\rm edge}(\rho,\sigma,\theta)=\frac{e^{i\theta(N-1/2)}}{\sqrt{2\pi N}} \frac{e^{-\rho^2-\sigma^2}}{2\pi i \sin \frac{\theta}{2}},
\end{equation}
which follows from a careful saddle point treatment of the incomplete gamma function. Setting $z=\sqrt{N}$, $w=\sqrt{N}e^{i\theta}$, the edge kernel may be simply interpreted as the chiral Dirac propagator $\propto \frac{1}{z-w}$ dressed by a gaussian enveloppe, so that the CFT regime holds within a distance $O(1)$ from the boundary.

Inserting (\ref{eq:edgekernel}) in (\ref{eq:c2d2}) and extending the integration over $\rho,\sigma$ to $\mathbb{R}$ yields a contribution
\begin{align}\nonumber
E(\theta)&=\int_{-\sqrt{N}}^\infty (\sqrt{N}+\rho)d\rho\int_{-\sqrt{N}}^\infty (\sqrt{N}+\sigma)d\sigma \,|V_{\rm edge}(\rho,\sigma,\theta)|^2\\
&=
 -\frac{1}{8\pi^2\sin^2\frac{\theta}{2}}.
\end{align}
in the limit $N\to\infty$. 
Integrating those contributions twice using the symmetry $\theta\to 2\pi-\theta$, and fixing the $\theta$-independent part using a different method explained in appendix \ref{app:C2}, we finally obtain
\begin{equation}
 \kappa(\theta)=\frac{\sqrt{N}}{\pi^{3/2}}+\frac{1}{2\pi^2}\log \left(\sqrt{N}\sin \frac{\theta}{2}\right)
 -\frac{1+(\pi-\theta)\cot \theta}{4\pi^2},
\end{equation}
up to a constant that does not depend on $\theta$, and subleading corrections in $1/\sqrt{N}$. In this result the first term is the area law contribution \cite{arealaw_proof}, the second is the CFT term, the third stems from the corner near the origin. We note that similar corner terms have been studied numerically in Ref.~\cite{RodriguezSierra2} for the entropy.

 \begin{figure*}[htbp]
    \begin{tikzpicture}[scale=0.85]
  \def\rad{0.8};
  \def\thet{75};
  \fill[white,inner color=red!80!white,outer color=red!50!white] (\rad,0) arc [radius=\rad,start angle=0,delta angle=\thet]
    -- ({3*cos(\thet)},{3*sin(\thet)}) arc [radius=3,start angle=\thet,delta angle=-\thet] -- cycle;
     \fill[blue,opacity=0.5] (0,0) -- (1.8cm,0mm) arc (0:360:1.8cm) -- (0,0);
      \draw[color=dred,ultra thick] (\rad,0) arc [radius=\rad,start angle=0,delta angle=\thet];
      \draw[color=dred,ultra thick] (\rad,0) -- (3,0);
      \draw[color=dred,ultra thick] ({\rad*cos(\thet)},{\rad*sin(\thet)}) -- ({3*cos(\thet)},{3*sin(\thet)});
    \draw[white,->,very thick] (1.3cm,0mm) arc (0:\thet:1.3cm);
    \draw[white] (1.2,1) node {\large{$\theta$}};
    \draw[very thick,dashed] (0,0) circle (1.8cm);
    \draw[->,very thick,black] (0,0) -- ({\rad*cos(45)},{\rad*sin(45)});
    \draw[->,very thick,black] (0,0) -- ({1.8*cos(37.5)},{-1.8*sin(37.5)});
    \draw (1,-0.45) node {$R$};
    \draw[black] (0.2,0.4) node {$r$};
     \draw (0,3.3) node {$S_n(r,\theta)$};
    \begin{scope}[xshift=8cm]
  \def\thet{180};

  \fill[white,inner color=red!80!white,outer color=red!50!white] (\rad,0) arc [radius=\rad,start angle=0,delta angle=\thet]
    -- ({3*cos(\thet)},{3*sin(\thet)}) arc [radius=3,start angle=\thet,delta angle=-\thet] -- cycle;
      \fill[blue,opacity=0.5] (0,0) -- (1.8cm,0mm) arc (0:360:1.8cm) -- (0,0);
    \draw[color=dred,ultra thick] (\rad,0) arc [radius=\rad,start angle=0,delta angle=\thet];
    \draw[color=dred,ultra thick] (\rad,0) -- (3,0);
    \draw[color=dred,ultra thick] ({\rad*cos(\thet)},{\rad*sin(\thet)}) -- ({3*cos(\thet)},{3*sin(\thet)});
    \draw[white,->,very thick] (1.3cm,0mm) arc (0:\thet:1.3cm);
    \draw[white] (0,1.5) node {\large{$\pi$}};
    \draw[very thick,dashed,opacity=0.6] (0,0) circle (1.8cm);
    \draw[->,very thick,black] (0,0) -- ({\rad*cos(45)},{\rad*sin(45)});
    \draw[->,very thick,black] (0,0) -- ({1.8*cos(37.5)},{-1.8*sin(37.5)});
    \draw (1,-0.45) node {$R$};
    \draw[black] (0.2,0.4) node {$r$};
    \draw (0,3.3) node {$S_n(r,\pi)$};
    \end{scope}
\begin{scope}[xshift=16cm]
   \def\thet{285};
   \fill[white,inner color=red!80!white,outer color=red!50!white] (\rad,0) arc [radius=\rad,start angle=0,delta angle=\thet]
    -- ({3*cos(\thet)},{3*sin(\thet)}) arc [radius=3,start angle=\thet,delta angle=-\thet] -- cycle;
    \fill[blue,opacity=0.5] (0,0) -- (1.8cm,0mm) arc (0:360:1.8cm) -- (0,0);
   \draw[color=dred,ultra thick] (\rad,0) arc [radius=\rad,start angle=0,delta angle=\thet];
   \draw[color=dred,ultra thick] (\rad,0) -- (3,0);
   \draw[color=dred,ultra thick] ({\rad*cos(\thet)},{\rad*sin(\thet)}) -- ({3*cos(\thet)},{3*sin(\thet)});
   \draw[white,->,very thick] (1.3cm,0mm) arc (0:\thet:1.3cm);
   \draw[white] (-0.4,-1.6) node {\large{$2\pi-\theta$}};
   \draw[very thick,dashed,opacity=0.6] (0,0) circle (1.8cm);
   \draw[->,very thick,black] (0,0) -- ({\rad*cos(45)},{\rad*sin(45)});
   \draw[->,very thick,black] (0,0) -- ({1.8*cos(37.5)},{-1.8*sin(37.5)});
   \draw (1,-0.45) node {$R$};
   \draw[black] (0.2,0.4) node {$r$};
   \draw (0,3.3) node {$S_n(r,2\pi-\theta)$};
   \end{scope}
 \end{tikzpicture}
 \vspace{-0.4cm}
  \caption{``Outer Camembert'' bipartition used to extract the chiral edge modes. We consider the entropy $S_n(r,\theta)$ of the region $\{z\in \mathbb{C}, |z|>r, 0<\arg z<\theta\}$, which is shown in shaded red. The droplet with radius $R=\sqrt{\alpha N}$ which is shown in blue for comparison. }
  \label{fig:camembert}
 \end{figure*}
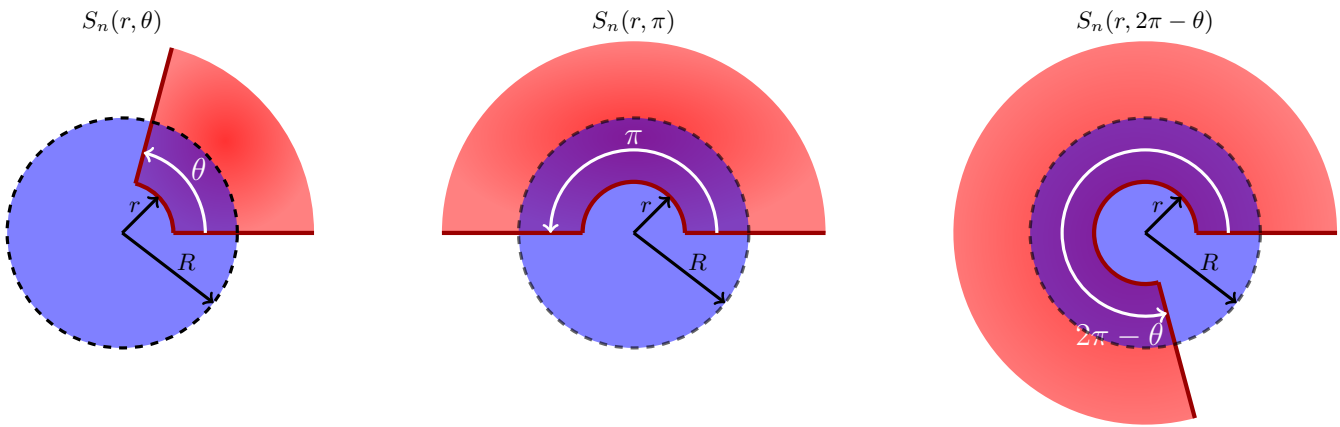

We finish with a remark. As already explained in the introduction, the $U(1)$ CFT is gaussian, which means there will be no CFT contributions to higher cumulants. This is not the case for the corner terms, however. We checked that such a $\theta-$dependent contribution is present for all cumulants, even though we were not able to find a general analytical formula. This contribution is not expected to be universal, and depends on the short distance specifics of the wave function considered.
\subsection{Extracting the chiral edge contributions}
The corner term we have computed in the previous section is an issue when trying to extract the edge CFT contribution to the entropy and fluctuations. To probe CFT contributions solely, we consider a different bipartition. Denote by $S_n(r,\theta)$ the entropy of the region $\{z\in \mathbb{C}, |z|>r, 0<\arg z<\theta\}$, shown in figure~\ref{fig:camembert}. Then, the linear combination
\begin{equation}\label{eq:camembertcombination}
 \Delta S_n(r,\theta)=S_n(r,\theta)+S_n(r,2\pi-\theta)-2S_n(r,\pi)
\end{equation}
allows to get rid of both the leading area law (since to the leading order $S_n \propto 2(R-r)+r\theta$) and the subleading corner piece, since the only relevant corner contribution corresponds to the same angle ($\pi/2$), and also gets cancelled in the linear combination. Therefore we expect only a constant piece, which is twice the entropy of the chiral edge,
\begin{equation}\label{eq:deltaS}
 \Delta S_n(r,\theta)=\frac{c}{6}\left(1+\frac{1}{n}\right) \log \sin \frac{\theta}{2}+o(N^0),
\end{equation}
where $c=1$ ($\bar{c}=0$). 
Equation (\ref{eq:deltaS}) holds provided both $r$ and $R-r$ are much larger than the correlation length. For example, any $r=aR$ for $a\in (0,1)$ and $R\to \infty$ works. We note  that an analogous scheme was designed in \cite{Crepel1}, for the cylinder geometry, where MPS techniques can be applied. See also \cite{KitaevPreskill,LevinWen} for previous variants aimed at extracting the topological EE. From our previous considerations, it is also clear that a similar result holds for the second cumulant of particle fluctuations, in which case
\begin{equation}\label{eq:deltakappa}
 \Delta \kappa (r,\theta)=\frac{K}{\pi^2}\log \sin \frac{\theta}{2}+o(N^0),
\end{equation}
consistent with the analytical result of the previous section. Here $K=1/\alpha=1$ is the Luttinger parameter. In fact, (\ref{eq:deltakappa}) can also be derived analytically using the method explained in section \ref{sec:disk_calculations}, even though we refrain from doing so here.

To confirm this idea numerically, we computed the linear combinations (\ref{eq:camembertcombination}) for both entropy and fluctuations.
At the free fermion point such calculations for large particle number are straightforward. In practice, we use the fact that the Fredholm determinant formula for the FCS can be recast as 
\begin{equation}\label{eq:fcsqhediscrete}
 \chi(\lambda,r,\theta)=\det_{0\leq k,l\leq N-1}\left(\delta_{kl}+[e^{\lambda}-1]M_{kl}\right)
\end{equation}
where
\begin{equation}
 M_{kl}=\frac{\Gamma(1+\frac{k+l}{2},r^2)}{\sqrt{k!l!}}\frac{\sin \frac{\theta(k-l)}{2}}{\pi(k-l)},
\end{equation}
using standard tricks (see e.g. Ref.~\cite{Klich_2006bis}). Hence numerically exact results for both fluctuations and entanglement can be obtained for large particle numbers. 

\begin{figure}[htbp]
 \includegraphics[width=0.48\textwidth]{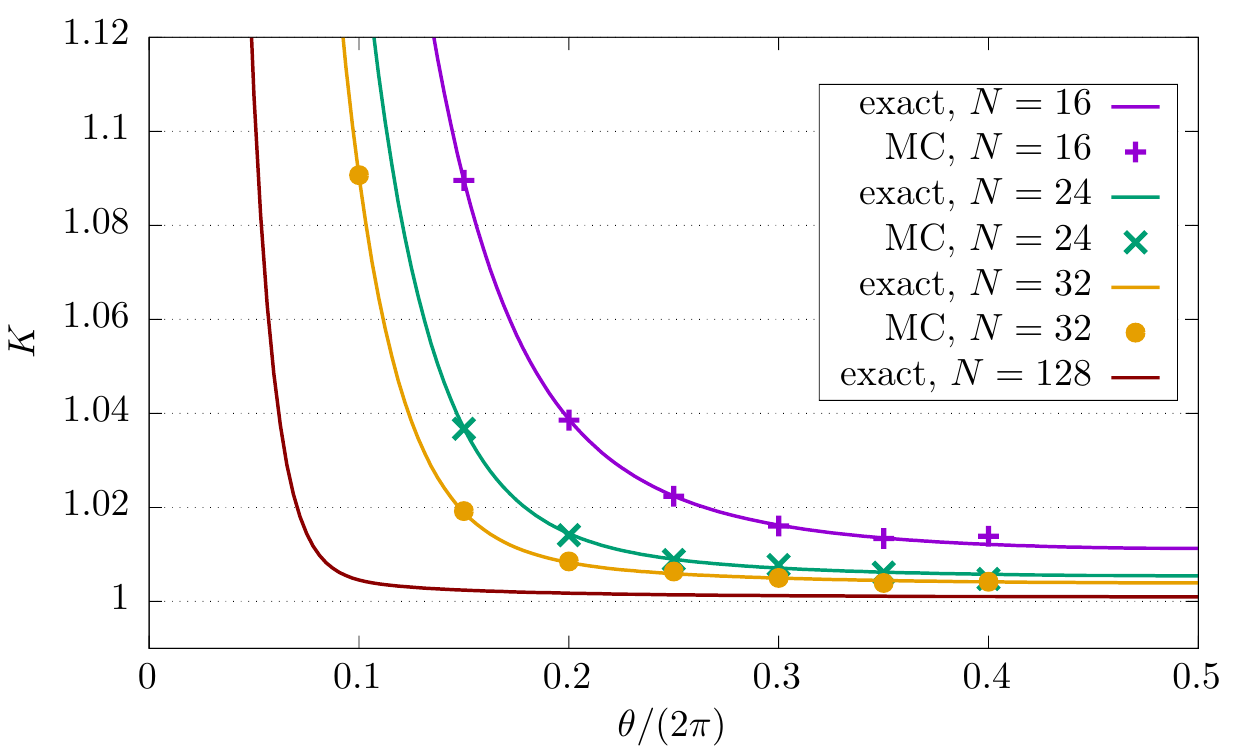}
 \caption{Extraction of the Luttinger parameter $K=1$ from the ratio $\frac{\pi^2\Delta \kappa(r,\theta)}{\log \sin\theta/2}$. The droplet has radius $R=\sqrt{N}$, and we choose $r=0.3R$ throughout. Thick lines are the exact free fermions results for $N=16,24,32,128$, while symbols represent Monte Carlo (MC) data. As can be seen the agreement is very good, and strongly suggests $K=1$. }
 \label{fig:C2_IQH}
\end{figure}

Let us first present the results for the second cumulants, which are shown in Fig.~\ref{fig:C2_IQH}.
We plot the ratio 
\begin{equation}
 \frac{\pi^2\Delta \kappa(r,\theta)}{\log \sin\theta/2}
\end{equation}
which should converge to $K=1$, in the limit $N\to \infty$. As can be seen, for not too large particle number the agreement is already fairly good. The numerical data for a larger particle number ($N=128$) shows impressive accuracy, and demonstrates the validity of the substraction procedure. 

\begin{figure}[htbp]
 \includegraphics[width=0.48\textwidth]{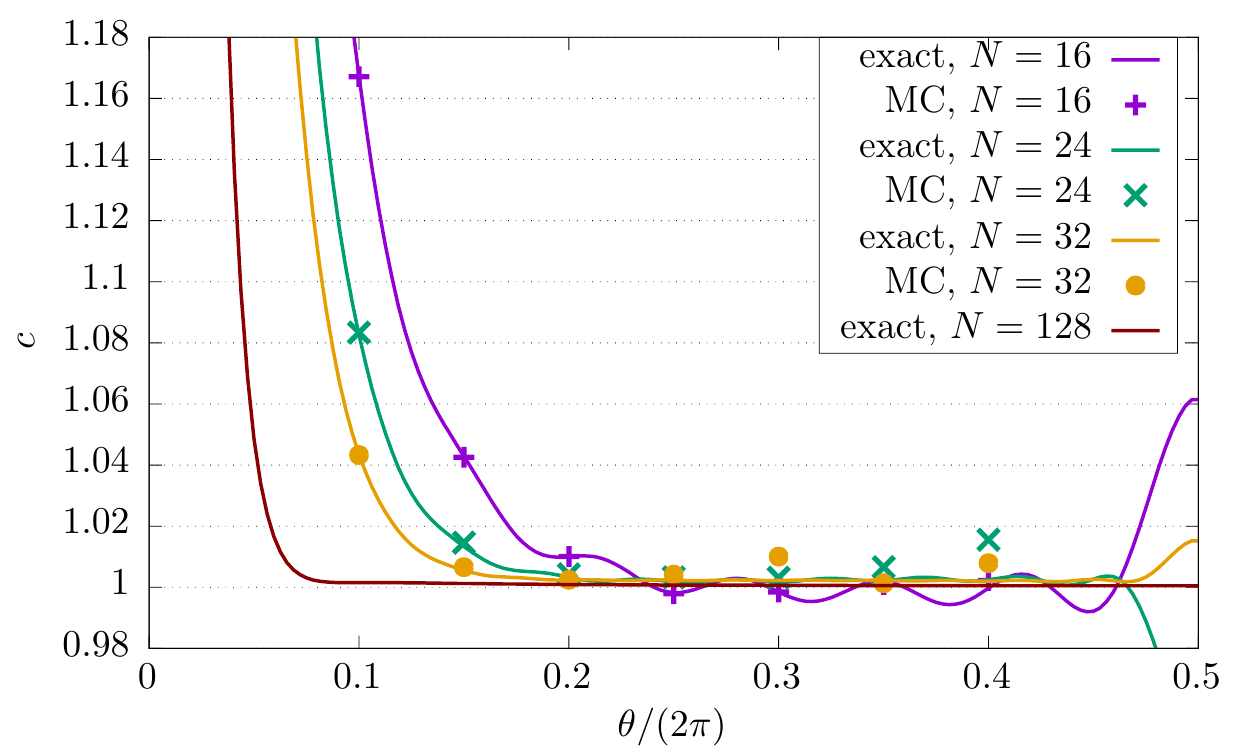}
 \caption{Extraction of the central charge $c=1$ from the ratio $\frac{4\Delta S_2(r,\theta)}{\log \sin\theta/2}$. The droplet has radius $R=\sqrt{N}$, and we choose $r=0.3R$ throughout. Thick lines are the exact free fermions results for $N=16,24,32,128$, while symbols represent Monte Carlo (MC) data. As can be seen the agreement is very good, and strongly suggests $c=1$.}
 \label{fig:S2_IQH}
\end{figure}

We then perform the same analysis for the second Renyi entropy $S_2$, see Fig.~\ref{fig:S2_IQH} for the numerical data. In this case we extract the central charge $c=1$, with very good agreement also. 
The choice of the second Renyi entropy is motivated by the fact that we will be able to compute it using Monte Carlo techniques even in the interacting case. However, we also checked that other entropies work very well too. Anticipating on future MC results we also show those, which match (up to to small error bars) the free fermion result perfectly. We note also the appearance of small oscillations in the second Renyi entropy. While such subleading corrections have been well studied for 1d quantum systems \cite{Cardy_2010}, a detailed investigation of those for QHE falls outside the scope of the present paper. However, we also observed that these corrections become bigger when increasing the Renyi index $n$, consistent with general 1d results.
\subsection{Numerical results for fractional quantum Hall states}
To demonstrate the broad validity of our approach also in the presence of interactions, we study the simplest interacting case, which is the bosonic Laughlin state with $\alpha=2$. There are no exact expressions for the fluctuations and entropy, so we have to rely on numerical simulations to probe the edge contributions. 

We do this using a Markov chain Monte Carlo algorithm. While accessing the fluctuations is a reasonably straightforward exercise, Renyi entropies can only be computed for integer index $n\geq 2$ using the swap method \cite{swap_ee}. In the following we focus on $n=2$. Let us write the wave function as 
\begin{equation}
\psi(z_1,\ldots,z_N)=\psi(z_i\in A|z_j\in B),
\end{equation}
namely, we (artificially for now) separate the particles in two sets depending on whether they are in the region $A$ or it complement $B=\mathbb{C}\backslash A$, and compute the wave function corresponding to the union of the two sets. Then (exponential minus the) second Renyi entropy may be expressed as 
\begin{align}\nonumber
 e^{-S_2}=\int_{\mathbb{C}^{2N}}& d^2z_1 \ldots d^2 z_N d^2w_1 \ldots d^2w_N\, \left|\psi(z_1,\ldots,z_N)\right|^2\\
 &\times |\psi(w_1,\ldots,w_N)|^2\, \textrm{Swap}_A (\mathbf{z},\mathbf{w}),
\end{align}
where
\begin{equation}
 \textrm{Swap}_A (\mathbf{z},\mathbf{w})=\frac{\psi(w_i \in A|z_j\in B)\psi(z_i\in A|w_j\in B)}{\psi(z_i \in A|z_j\in B)\psi(w_i\in A|w_j\in B)}
\end{equation}
swaps the particle configurations between the two copies $\mathbf{z}=\{z_1,\ldots,z_N\}$ and $\mathbf{w}=\{w_1,\ldots,w_N\}$, but only in subsystem $A$. Hence $e^{-S_2}$ can be computed by sampling two independent copies of FQH using a Metropolis-Hastings algorithm, and evaluating the swap operator. Even though the expectation value is a number in $(0,1)$, each realization of the swap is a complex number. 
Hence the method suffers from a ``sign'' problem, which can be mitigated using the tricks explained in Refs.~\cite{sign_trick,McMinisTubman,ee_composite}. In practice, the entropies can be computed to good precision for $N \lesssim 50$.

\begin{figure}[htbp]
\includegraphics[width=0.48\textwidth]{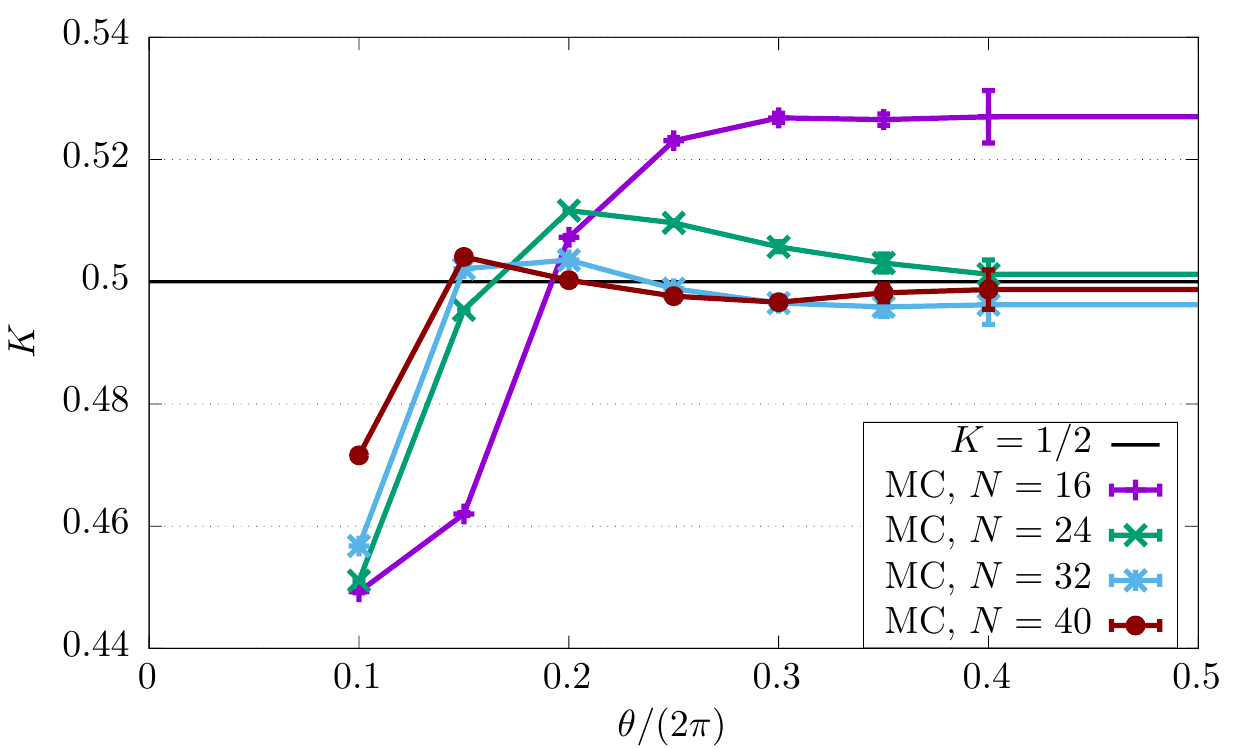}
 \caption{Extraction of the Luttinger parameter $K=1/2$ from the ratio $\frac{\pi^2\Delta \kappa(r,\theta)}{\log \sin\theta/2}$. Same procedure as for IQH. The symbols represent Monte Carlo (MC) data with error bars. As can be seen the agreement is very good, and strongly suggests $K=1/2$. Note that error bar increases significantly when $\theta$ becomes close to $\pi$. }
 \label{fig:C2_FQH}
\end{figure}

The results for the fluctuations are shown in Fig. \ref{fig:C2_FQH}. As can be seen the agreement with the expected result $K=1/2$ is very good, and improves further when increasing $N$.  
Let us make two important remarks when looking at the data. The CFT prediction is only valid in the long distance limit, which means the conformal scaling is not expected to hold when $\theta$ is close to a multiple of $\pi$. This explains why substantial deviations are still observed near $\theta=0$, even though we of course  expect 
$\pi^2 \Delta \kappa(r,\theta)/\log \sin \frac{\theta}{2}=K=1/2$ in the limit $N\to \infty$ for any $\theta\neq 0 \,\textrm{mod}\,\pi$. Another observation is that error bars increase significantly when $\theta$ is close to $\pi$. This is because $\Delta \kappa$ behaves as $\Delta \kappa(r,\theta)\sim -K\frac{(\theta-\pi)^2}{8\pi^2}$ for $\theta$ close to $\pi$. Therefore, it becomes necessary to compute each $\kappa(r,\theta)$ to very high accuracy to get a reliable estimate of $\Delta\kappa(r,\theta)$, and extracting $K$ for such angles requires extremely long simulations, even for small particle numbers. In practice, we did not go further than $\theta/(2\pi)=0.4$.

\begin{figure}[htbp]
 \includegraphics[width=0.48\textwidth]{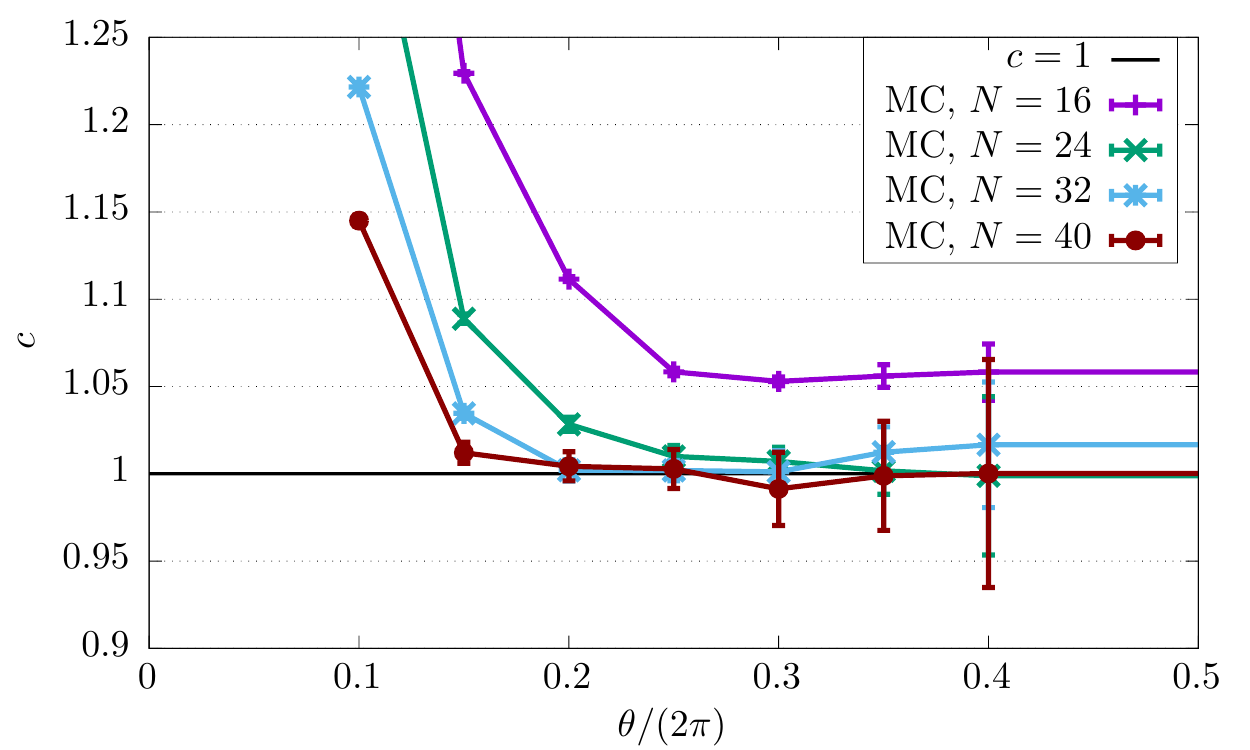}
 \caption{Extraction of the central charge $c=1$ from the ratio $\frac{4\Delta S_2(r,\theta)}{\log \sin\theta/2}$. The symbols represent Monte Carlo (MC) data. As can be seen the agreement is very good, and strongly suggests $c=1$. }
 \label{fig:S2_FQH}
\end{figure}

The entropy is expected to behave differently, since the central charge is still one in that case. As can be seen in Fig. \ref{fig:S2_FQH}, the data strongly suggests $c=1$, with good agreement already for a modest number of particles. This is further confirmation that entanglement and fluctuations probe different universal data in the underlying chiral CFT. The two comments regarding numerical accuracy for the fluctuations also apply to the entropy. Nevertheless, the final accuracy to which we obtain the central charge is quite remarkable: over a wide range of values of $\theta$, the error on $c$ is smaller than a percent.
\section{Conclusion}
In this paper, we have studied the entanglement and fluctuation scaling of the simplest quantum Hall wave functions. We investigated geometries where an explicit logarithmic contribution from the chiral edge modes can be calculated, in addition to the more familiar area law term. This contribution is exactly that predicted by general conformal field theory arguments, with central charge $c=1$, $\bar{c}=0$. The result differs from well known results in gapless spin chains, for which holomorphic and antiholomorphic contributions cannot be separated ($c=\bar{c}$). Here this is possible due to the fact that the system is two-dimensional, with a gapped bulk. Most of our analysis focused on integer quantum Hall wave function for which complete analytical calculations are possible, starting from the microscopic wave function. We also checked this idea to remarkable numerical precision in the original Laughlin wave function. This was done using Monte Carlo techniques, complementing the MPS numerical results of \cite{PhysRevB.88.155314,Crepel1,Crepel2,Crepel3} in cylinder geometries. We have also found a simple generalization to the multi-interval case in the non-interacting case.

The present work opens up several interesting directions for further research, some of which we list below. The first would be to try and study other model states constructed out of CFT correlators \cite{MooreReadstate,ReadRezayistate}. An example is the Moore Read state, for which we expect even more different behavior between fluctuations and entropy. Indeed the central charge, $c=1+1/2=3/2$ will be the sum of the bosonic and fermion sectors, while we expected fluctuations to be only sensitive to the charge (bosonic) sector. This example can still be treated by Monte Carlo techniques (the bottleneck in the update involves computing a Pfaffian, which can be done in complexity $N^3$), and possibly analytical ones too, since it maps to a two component gas, but still with logarithmic interactions. 

Back to the Laughlin and IQH states, another possibility is to look at the entanglement spectrum in geometries with chiral boundary, and see whether one can identify features specific to the bulk and edge just by looking at the spectrum. It would be also interesting to investigate other droplet shapes, as e. g. in Ref. \cite{DiFrancesco}. 

On the mathematical side, there are known connections to the theory of random matrices (where e.g. the Laughlin IQH maps to the Ginibre ensemble of RMT), or, more generally, two-dimensional Coulomb gases with logarithmic interactions, which can be treated using probabilistic techniques. For example, it is possible to prove a central limit theorem (see \cite{RiderVirag} for $\alpha=1$, and e.g. \cite{Leble_Serfaty,Bauerschmidt} for general $\alpha$) for linear statistic associated to smooth test functions. In this case the gaussian fluctuations are of order one. The particle statistics we looked at correspond to characteristic functions of a given region in this context, so are not smooth. For this reason the results presented here fall outside the scope of known theorems. While we were able to prove an analogous result in the strip geometry, the disk geometry seems to be more complicated.  However, we do believe that a proof of central limit theorem for particle fluctuations is possible in this case also, and plan to come back to such problems in the future. Another more long term goal would be to use Coulomb gas techniques to prove the topological entropy result \cite{KitaevPreskill,LevinWen} in the Laughlin state. 

Finally, perhaps the most intriguing questions lie in higher dimensional problems. While it is well known that Quantum Hall physics can occur in any (even) dimension, not much is known regarding particle fluctuations and entanglement, and whether they reveal universal features of the underlying physics.

\acknowledgments
We are grateful to J\'er\^ome Dubail, C\'ecile Repellin and Laurent Charles for early collaboration on this topic. We also thank Nicolas Regnault, Valentin Cr\'epel, Pasquale Calabrese and Semyon Klevtsov for discussions. Benoit Estienne was supported by the grant ANR TopO No. ANR-17-CE30-0013-01. 

Shortly before completion of the present paper, we became aware of a related work \cite{Witczak}.

\pagebreak
\appendix
\onecolumngrid

\section{Proof of gaussianity in the half-plane geometry}
\label{app:C0}
Most of our analysis relied on the following physically well known but crucial  fact: the $U(1)$ chiral CFT describing the edge is free, so has, by definition, gaussian fluctuations. For this reason, we expect that the (edge contribution of the) higher cumulants are suppressed with respect to variance in the semiclassical $l_B\to 0$ limit. Then, using the known relation between cumulants and entropy, the leading logarithmic term of the entanglement entropy can be reconstructed from the variance only, for which we derived an exact asymptotic expansion.

It is in general more difficult to obtain an asymptotic expansion of the full counting statistics (or equivalently, entanglement entropies) from the microscopic model, and prove gaussianity of the fluctuations. Fortunately, we were able to do this for a single interval in the half-plane geometry, we explain the method here. The appendix is rather long, so we start with a summary of the main steps and results. 

We denote by $\chi(\lambda)$ the generating function for the moments of particle fluctuations in region $A$ which we take to be $\mathbb{R}\times [-l/2,l/2]$ (see the main text), also known as full counting statistics. Its logarithm generates cumulants, $\log \chi(\lambda)=\sum_{p\geq 1}\kappa_p \lambda^p/p!$. As is well known, for free fermions it is given by a Fredholm determinant involving the kernel $K$ mentioned in the main text. After further manipulations described in appendix \ref{app:recast}, it may be recast under the form
\begin{equation}\label{eq:thefreddet}
 \chi(\lambda)=\det(I-\omega V)\qquad,\qquad \omega=1-e^\lambda,
\end{equation}
where the kernel $V$ acts on $L^2(R)$ with $R$ the region $R=[\frac{a}{l_B},\frac{b}{l_B}]$. It takes the form $V(x,y)=f(x-y)$, with
\begin{equation}\label{eq:thekernel}
 f(x)=\frac{1}{2 \pi} \frac{l}{l_B} \textrm{sinc} \left(\frac{lx}{2l_B}\right) e^{-x^2/4}.
\end{equation}
Note that we use the letter $x$ here to match standard conventions on such kernels, even though it corresponds physically to momentum along the boundary of $A$. In the main text $x$ corresponds to $q/l_B$.
For the Fredholm determinant we use the definition
\begin{align}
 \det(I-\omega V)&=\exp\left(-\sum_{p=1}^\infty \frac{\omega^p}{p}\int_{R^p}dx_1\ldots dx_p V(x_1,x_2)V(x_2,x_3)\ldots V(x_p,x_1)\right).
\end{align}
The fact that the kernel only depends on the difference of argument helps greatly in our analysis. Such operators are dubbed Toeplitz or Wiener-Hopf operators, and their determinants have been widely studied. We need one more piece of notation before stating the results. For any sufficiently smooth function $g$ on $\mathbb{R}$, we note $\hat{g}$ its Fourier transform, using the following conventions
\begin{equation}
 \hat{g}(k)=\int_\mathbb{R}dx e^{-ikx}g(x)\qquad,\qquad g(x)=\int_\mathbb{R}\frac{dk}{2\pi}e^{ikx} \hat{g}(k),
\end{equation}
for the transform and its inverse. For example,
\begin{equation}\label{eq:oursymbol}
 \hat{f}(k)=\textrm{er}\left(k - \frac{l}{2l_B}\right) -\textrm{er}\left(k + \frac{l}{2l_B}\right)
\end{equation}
is the Fourier transform of the kernel (\ref{eq:thekernel}), where $\textrm{er}(x)=\int_x^\infty \frac{dt}{\sqrt{\pi}}e^{-t^2}$ is an error function.
In appendix~\ref{app:Kac}, we establish the crucial identity
\begin{equation}\label{eq:magnificient}
\log  \det(I-\omega V)=\frac{b-a}{l_B}L(0)+\int_0^\infty dx\,x L(x)^2+O\left(e^{-1/l_B}\right),
\end{equation}
where 
\begin{equation}
\hat{L}(k)=\log(1-\omega \hat{f}(k)), 
\end{equation}
 and $L(x)$ is the inverse Fourier transform of $\hat{L}(k)$. \\
 
 It is important to realize that the two-terms in the \emph{r.h.s.} of \eqref{eq:magnificient}, namely $\frac{b-a}{l_B}L(0)$ and $\int_0^\infty dx\,x L(x)^2$, correspond to bulk and edge contributions, respectively. The first one is proportional to the width of the strip $b-a$, and besides a trivial volume term it does not depend on $l$, while the second one does not depend on $b-a$, but purely on $l/l_B$.  \\

 The reader well-acquainted with the literature on operator determinants may have spotted that the equation (\ref{eq:magnificient}) is formally identical to a famous result of Kac \cite{kac1954}, itself a continuum analog of the celebrated strong Szeg\"o limit theorem \cite{Szego}. The theorem considers a Toeplitz operator with kernel given once and for all, and then a limit where the size of the interval $R$ diverges. Under smoothness assumptions, both $L(0)$ and the integral on the rhs are finite, which means the logarithm of the determinant grows linearly with the size of the interval, up to a constant that is determined exactly. We will see that this is not the case for us.

For our determinant we consider a limit $l_B\to 0$, which does imply that the interval length diverges as in Kac's theorem. However the big difference is that the kernel $V$ does itself depend on $l_B$ in our case. Even worse, it actually becomes singular in the limit $l_B\to 0$. This makes it impossible to apply Kac's result, and a new analysis is required. This is done in appendix~\ref{app:Kac}, were we establish formula (\ref{eq:magnificient}). 
Curiously, most of the steps in Kac's proof can still be used to derive (\ref{eq:magnificient}), even though the order of magnitude of many terms is completely different in our problem.  

The last step is to analyze the asymptotic behavior of both contributions in (\ref{eq:magnificient}), in the limit $l_B\to 0$. This is done in appendix \ref{app:justcompute}. We obtain
\begin{align}
\log  \det(I-\omega V)& =\frac{(b-a)}{2\pi l_B} \frac{l}{l_B}\log(1-\omega)+\frac{b-a}{l_B} G_+(\omega,0) + \frac{\log(1-\omega)^2}{2} \Delta \kappa_2+ A_0(\omega) + A_1(\omega) \frac{l_B}{l} + O(l_B^2) 
\label{eq:almostfinalexpansion}
\end{align}
where $\Delta \kappa_2$ is (twice) the edge variance of the main text, namely
\begin{align}
\Delta \kappa_2 = \frac{l^2}{2\pi^2 l_B^2}\, _2F_2\left(1,1;\frac{3}{2},2;-\frac{l^2}{2l_B^2}\right) = \frac{1}{\pi^2} \log \frac{l}{l_B}  + \frac{\log 2+\gamma}{2 \pi ^2} + O(l_B^2), \qquad (l_B \to 0)
\end{align}
and  $A_0(\omega)$ and $A_1(\omega)$ are given by
\begin{align}
A_0(\omega)   = - \frac{\log(1-\omega)}{\pi} \int_{0}^{\infty} dx \, G_-(\omega,x) e^{-\frac{x^2}{4}}  +  \int_{0}^{\infty} dx \, x\frac{G^2_-(\omega,x) +G^2_+(\omega,x)}{2}\qquad, \qquad A_1(\omega)  = \frac{\log(1-\omega)}{\pi} G_+(\omega,0),
\end{align}
and the functions $G_+(\omega,x)$ and $G_-(\omega,x)$ are given by
\begin{align}
G_+(\omega,x)  & = \int_\mathbb{R}\frac{dk}{\pi}\cos \left(kx\right) \left[\log(1-\omega \textrm{er}(k))- \log(1-\omega)\textrm{er}(k)\right], \\
G_-(\omega,x) &= \int_\mathbb{R}\frac{dk}{\pi}\sin \left(kx\right)\left[\log(1-\omega \textrm{er}(k))- \log(1-\omega)\textrm{er}(k)\right]  \,.
\end{align}
where recall $\textrm{er}(x)=\int_x^\infty \frac{dt}{\sqrt{\pi}}e^{-t^2}$ is an error function. 
 In the result \eqref{eq:almostfinalexpansion}, the first two terms come from $L(0)$, while the logarithmic divergence and the subleading polynomial corrections comes from the integral in (\ref{eq:magnificient}). 

Recalling $\omega=1-e^\lambda$, the final result for the FCS reads
\begin{equation}\label{eq:finalexpansion}
 \log \chi(\lambda)=\frac{(b-a)}{2\pi l_B} \frac{l}{l_B}\lambda+\frac{b-a}{l_B} G_+(1-e^{\lambda},0) +\frac{\lambda^2}{2}\Delta \kappa_2+ A_0(1-e^{\lambda}) + A_1(1-e^{\lambda}) \frac{l_B}{l} + O(l_B^2) 
\end{equation}
where $\Delta \kappa_2 \sim \frac{1}{\pi^2} \log \frac{l}{l_B}$ is the the only term containing a log divergence. In this formula, the first term is a bulk term, it simply states that the mean number of particle is proportional to the ``volume''. This is a simple consequence of the incompressibility of the QH droplet, which implies a uniform density $(2\pi l_B^2)^{-1}$. The second, proportional to $1/l_B$ is the area law contribution of Ref.~\cite{arealaw_proof}. The logarithmic term is the edge CFT contribution, and is new. Importantly it is proportional to $\lambda^2$, which means it affects only the second cumulant.  The terms $A_0$ only contributes to the even cumulants $\kappa_n$ with $n \geq 4$, while $A_1$ only contributes to the odd cumulants $\kappa_n$ with $n \geq 3$. So at leading order we find the following edge contributions to the FCS : 
\begin{align}
\boxed{\Delta\kappa_2 \sim \frac{1}{\pi^2} \log \frac{l}{l_B}, \qquad \Delta\kappa_{2n} = O(1), \qquad \Delta\kappa_{2n-1} = O(l_B), \qquad ( n \geq 2)}
\end{align}
where $\Delta \kappa_n$ denote the edge contribution to the n${}^{\textrm{th}}$ cumulant of the number of particles in region $A$. Note that the variance is twice the one of the main text, because the geometry considered in this appendix has two edges.

Let us finally comment on the relation to the entanglement entropy. It can be deduced \cite{Klich_2009} from the FCS, as explained in the text. However, a more transparent approach is to establish, in the spirit of \cite{arealaw_proof}, the slightly more general result
\begin{equation}\label{eq:themostgeneral}
 \textrm{Tr}\, h(V)=\frac{(b-a)}{2\pi l_B} \frac{l}{l_B} h(1)+\frac{b-a}{\pi l_B}\int_{\mathbb{R}}dx\left[h( \textrm{er}(x))- h(1)\textrm{er}(x)\right]
 +\frac{1}{\pi^2}\left[\,\int_0^1 \frac{h(z)-zh(1)}{z(1-z)}dz\,\right]\log \frac{l}{l_B}+O(l_B^0),
\end{equation}
which follows easily from our method. Here $h$ is any sufficiently smooth function on $[0,1]$ satisfying $h(0)=0$. Now, the choice $h(z)=\log(1+(e^\lambda-1) z)$ gives back the FCS, while the choice $h(z)=\frac{\log\left[z^n+(1-z)^n\right]}{1-n}$ gives the R\'enyi entropies. In the latter case the bulk term is absent since $h(1)=0$, as expected. The coefficient of the logarithm can be explicitely computed,
\begin{equation}
 \frac{1}{\pi^2}\int_0^1 \frac{h(z)-zh(1)}{z(1-z)}dz=\frac{1}{6}\left(1+\frac{1}{n}\right),
\end{equation}
and we recover the well-known conformal dependence of R\'enyi entropies on the R\'enyi index $n$. 
\subsection{Fredholm-Toeplitz determinant representation of the FCS}
\label{app:recast}
In this section we establish (\ref{eq:thefreddet}). Our starting point is the Fredholm determinant one gets starting from the two-dimensional problem
\begin{equation}
 \chi(\lambda)=\det(I-\omega K),
\end{equation}
where the kernel is given by (see equation \eqref{strip_kernel} in the main text)
\begin{equation}
 K(x_1,y_1;x_2,y_2)=\frac{1}{2\pi\sqrt{\pi} l^3_B} e^{- \frac{x_1^2 + x_2^2}{2l_B^2}} \int_{a}^{b} dq \, e^{\frac{q (z_1+\bar{z}_2)}{l_B^2}}   e^{-\frac{q^2}{l_B^2}} ,
\end{equation}
where $z_j = x_j + i y_j$. The integrations are performed on the domain $A=\mathbb{R}\times \mathcal{A}$. Let us now compute all traces of powers of the 2d kernel. A simplification occurs since all the integrals on $x_j$ are gaussian, so can be performed exactly. Integrating over $x$ yields
\begin{align}
\textrm{Tr}\, K^p  &   =\int d^px d^py \, K(x_1,y_1;x_2,y_2)K(x_2,y_2;x_3,y_3)\ldots K(x_p,y_p;x_1,y_1)\\
 & = \left(\frac{1}{2\pi l_B^2} \right)^n \int_{\mathcal{A}^p} d^p y \,  \int_{[a,b]^p} d^pq \,  e^{i \sum_{j=1}^p \frac{q_j (y_j - y_{j+1})}{l_B^2}}    e^{- \sum_{j=1}^p \frac{(q_j-q_{j-1})^2}{4l_B^2}}
\end{align}
where $y_{p+1} = y_1$, or equivalently
\begin{align}
\textrm{Tr}\, K^p  =  \left(\frac{1}{2\pi l_B^2} \right)^p \int_{\mathcal{A}^p} d^py \,  \int_{[a,b]^n} d^pq \,  e^{i \sum_{j=1}^p \frac{y_j( q_j - q_{j-1})}{l_B^2}}    e^{- \sum_{j=1}^p \frac{(q_j-q_{j-1})^2}{4l_B^2}}, 
\end{align}
with $q_0 = q_p$. 
We get a Fredholm determinant with kernel 
\begin{align}
 \tilde{V}(q,q')&=\int_\mathcal{A} \frac{dy}{2\pi l_B^2} e^{-\frac{(q-q')^2+4i y(q-q')}{4l_B^2}}
 \end{align}
In  particular when $\mathcal{A}$ is a single interval, say $\mathcal{A} = \left[ - \frac{l}{2} , \frac{l}{2} \right]$, this yields
\begin{align}
 \tilde{V}(q_1,q_2)&=\frac{\sin \frac{l(q_1-q_2)}{2l_B^2}}{\pi(q_1-q_2)}e^{-\frac{(q_1-q_2)^2}{4l_B^2}}.
\end{align}
The integrations are performed on the interval $[a,b]$ (with the obvious continuation for $q_1=q_2$). Finally, a simple change of variable $q_j=l_B x_j$ yields (\ref{eq:thefreddet}) on the interval $[a/l_B,b/l_B]$, as advertized.

\subsection{A Szego-Kac type result}
\label{app:Kac}
In this section, we study the determinant (\ref{eq:thefreddet}), by a brute force trace expansion, following the approach of \cite{kac1954} almost to the letter.  Without loss of generality we set the interval of integration to be $R=[0,t]$ where $t=(b-a)/l_B$. The crucial ingredient that makes the present analysis work is that the kernel (\ref{eq:thekernel}) decays exponentially fast with distance, for any $l_B$. Now, let us consider an integer $p\geq 3$, and compute
\begin{equation}
 \textrm{Tr}\, V^p=\int_{R^p} dx_1\ldots dx_p f(x_1-x_2)f(x_2-x_3)\ldots f(x_p-x_1).
\end{equation}
Introduce $\chi$, the characteristic function of the interval $R$. $\chi(x)=1$ if $x\in [0,t]$, $\chi(x)=0$ otherwise. Then, the trace may be rewritten as
\begin{equation}
 \textrm{Tr}\, V^p=\int_{\mathbb{R}^p} dx_1\ldots dx_p \chi(x_1)\ldots \chi(x_p) f(x_1-x_2)f(x_2-x_3)\ldots f(x_p-x_1).
\end{equation}
Next, we make the change of variable $y_1=x_1$, $y_2=x_2-x_1$, \ldots, $y_p=x_p-x_{p-1}$. The Jacobian is one, so
\begin{equation}\label{eq:somechitrace}
  \textrm{Tr}\, V^p=\int_{\mathbb{R}^p} dy_1\ldots dy_p \chi(y_1)\chi(y_1+y_2)\ldots \chi(y_1+\ldots+y_p)f(-y_2)\ldots f(-y_p)f(y_2+\ldots+y_p).
\end{equation}
The integral on $y_1$ can be performed using the identity
\begin{equation}
 \int_\mathbb{R}dy_1\chi(y_1)\chi(y_1+y_2)\ldots \chi(y_1+\ldots+y_p)=\max\Big(0,t-\max(0,y_2,\ldots,y_2+\ldots+y_q)+\min(0,y_2,\ldots,y_2+\ldots+y_p)\Big)
\end{equation}
Looking at (\ref{eq:somechitrace}), it is easy to see that relaxing the outer $\max(0,\ldots)$ constraint amounts to neglecting terms which are exponentially small in $(b-a)^2/l_B^2$, since such terms correspond to $\max(x_i) - \min(x_i) \geq t$ in the original variables. Therefore, we have, up to exponentially small terms,
\begin{align}
 \textrm{Tr}\, V^p&=t\int_{\mathbb{R}^{p-1}} dy_2\ldots dy_p f(-y_2)\ldots f(-y_p)f(y_2+\ldots+y_p) \\ 
  & -2 \int_{\mathbb{R}^{p-1}} dy_2\ldots dy_p \max(0,y_2,\ldots,y_2+\ldots +y_p) f(-y_2)\ldots f(-y_p)f(y_2+\ldots+y_p)
   \end{align}
   where we used the fact that $f$ is even in the last line. The above may be rewritten as
   \begin{align}
  \textrm{Tr}\, V^p&=t \int_\mathbb{R}\frac{dk}{2\pi}\hat{f}(k)^p-2\int_{\mathbb{R}^{p-1}} dy_2\ldots dy_p \max(0,y_2,\ldots,y_2+\ldots +y_p) f(-y_2)\ldots f(-y_p)f(y_2+\ldots+y_p).
 \end{align}
 For the first term, we have recognized a simpler expression in terms of the Fourier transform, $\hat{f}$. Both integrals converge, but the second one on the rhs looks much more complicated. Nevertheless, it can also be expressed in terms of $\hat{f}$, as follows
 \begin{equation}
  \int_{\mathbb{R}^{p-1}} dy_2\ldots dy_p \max(0,y_2,\ldots,y_2+\ldots +y_p)f(-y_2)\ldots f(-y_p)f(y_2+\ldots+y_p)=\frac{p}{2}\int_0^\infty dx\, x \sum_{j=1}^{p-1} \frac{f_j(x)f_{n-j}(x)}{j(p-j)}
 \end{equation}
 where $f_j(x)$ is the inverse Fourier transform of $(\hat{f})^j$. 
We refer to \cite{kac1954} for a proof of this remarkable identity. Using this, we obtain
\begin{equation}\label{eq:importantintermediate}
  \textrm{Tr}\, V^p=t\int_\mathbb{R}\frac{dk}{2\pi}\hat{f}(k)^p-p\int_0^\infty dx\, x \sum_{j=1}^{p-1} \frac{f_j(x)f_{p-j}(x)}{j(p-j)}
\end{equation}
up to exponentially small corrections. Obtaining a uniform bound in $p$ requires only little extra work, and resumming everything we finally recognise (\ref{eq:magnificient}).
\subsection{Final pieces of asymptotic analysis}
\label{app:justcompute}
Our final task is to perform an asymptotic analysis of the two leading terms in (\ref{eq:magnificient}). Both involve the function $L(x)$, defined as 
\begin{equation}
 L(x)=\int_{\mathbb{R}} \frac{dk}{2\pi} e^{ikx}\log(1-\omega \hat{f}(k)),
\end{equation}
where $\hat{f}(k)$ is given by (\ref{eq:oursymbol}). Let us start with the first, which is $L(0)$. Intuitively for large $l_B$, $\hat{f}$ is close to the rectangular function $\textrm{rect}(k)$ which evaluates to one for $k\in [-l/2l_B,l/2l_B]$ and zero otherwise, similar to a Fermi sea. Based on this intuition, and since $\log(1-\omega \,\textrm{rect}(k))=\log(1-\omega)\textrm{rect}(k)$, we expand 
\begin{equation}
\log(1-\omega \hat{f}(k)) = \log(1-\omega) \hat{f}(k) + \left[ \log(1-\omega \hat{f}(k))-\log(1-\omega) \hat{f}(k)  \right]
\end{equation}
yielding
\begin{equation}
 L(0)=\frac{l}{2\pi l_B}\log(1-\omega)+\int_\mathbb{R} \frac{dk}{2\pi}\left[\log(1-\omega \hat{f}(k))-\log(1-\omega)\hat{f}(k)\right],
\end{equation}
The first term int the \emph{r.h.s.} is the extensive bulk law in (\ref{eq:almostfinalexpansion}), while the integral is dominated by neighborhoods of $k = \pm l/2l_B$, and up to exponentially small terms as $l_B \to 0$, is equal to
\begin{equation}
\int_\mathbb{R} \frac{dk}{2\pi}\left[\log(1-\omega \hat{f}(k))-\log(1-\omega)\hat{f}(k)\right] = \int_\mathbb{R} \frac{dk}{\pi}F(\omega,k)
\end{equation}
where
\begin{equation}
 F(\omega,k)=\log(1-\omega \textrm{er}(k))-\log(1-\omega) \textrm{er}(k),
\end{equation}
and we recover the area law contribution to (\ref{eq:almostfinalexpansion}).
Let us now come to the most interesting one, which is the integral that contains the edge contribution, namely
\begin{equation}
 \int_0^\infty dx\,x L(x)^2. 
\end{equation}
To study it, we use a similar method as above. We decompose
\begin{align}
 L(x) & = \log(1-\omega)f(x)+ G(\omega,x) 
 \end{align}
 where
 \begin{align}
 G(\omega,x) &= \int_\mathbb{R}\frac{dk}{\pi}\cos \left(kx + \frac{l}{2l_B}x\right) F(\omega,k) =  \cos \left(\frac{l}{2l_B}x\right)  G_{+}(\omega,x)- \sin \left(  \frac{l}{2l_B}x\right) G_{-}(\omega,x)
   \end{align}
The functions $G_+(\omega,x)$ and $G_-(\omega,x)$ are given by
\begin{align}
G_+(\omega,x)  = \int_\mathbb{R}\frac{dk}{\pi}\cos \left(kx\right) F(\omega,k) , \qquad  G_-(\omega,x) &= \int_\mathbb{R}\frac{dk}{\pi}\sin \left(kx\right)F(\omega,k) \,.
\end{align}
and they respectively even (odd) under $\lambda \to - \lambda$. This follows from the fact that $F(\omega,k) $ is invariant under $\omega \to \omega/(\omega-1)$ (corresponding to $\lambda \to -\lambda$) and $k \to -k$.  Using this decomposition, we obtain
\begin{align}
 \int_0^\infty dx\,x L(x)^2 &=\log(1-\omega)^2\int_0^\infty dx\,x f(x)^2+2\log(1-\omega)\int_0^\infty dx\,x f(x) G(\omega,x) +\int_0^\infty dx\,x  G(\omega,x)^2\label{eq:fullasymptotic}
  \end{align}
Above, the first term coincides with (twice) the explicit result explained the main text, see Eq.~(\ref{strip_plane_exact}).
  \begin{align}
\int_0^\infty dx\,x f(x)^2 &=\frac{l^2}{4\pi^2 l_B^2}\, _2F_2\left(1,1;\frac{3}{2},2;-\frac{l^2}{2l_B^2}\right) = \frac{1}{2\pi^2} \log \frac{l}{l_B}  + O(1), \qquad (l_B \to 0)
\end{align}
We recover the log divergence of (\ref{eq:almostfinalexpansion}), which is proportional to $\log(1-\omega)^2=\lambda^2$, thus proving that the cumulants $\kappa_n$ have no log divergence for $n \geq 3$. We note that (\ref{eq:fullasymptotic}) allows to obtain the asymptotic expansion to any polynomial order in $l_B$.  A closer look at the second and third terms reveals that the edge contribution $ \int_0^\infty dx\,x L(x)^2 $ to the cumulants  $\kappa_n$ is $O(1)$ for $n$ even and $O(l_B)$ for $n$ odd for $n \geq 3$. For instance the edge contribution to $\kappa_3$ is
   \begin{align}
\Delta \kappa_{3} \sim \frac{12}{(2\pi)^{5/2}} \frac{l_B}{l}  
    \end{align}
Indeed the asymptotic behavior of the the second and third terms is easily obtained as follow. The second one yields
\begin{align}
 \int_0^\infty dx f(x) G(x) & \simeq \frac{1}{\pi} \int_0^\infty dx   \sin \left(\frac{l x}{2l_B}\right)  e^{-x^2/4} \left( \cos \left(\frac{l}{2l_B}x\right)  G_{+}(x)- \sin \left(  \frac{l}{2l_B}x\right) G_{-}(x)\right) \\
  &   = -  \frac{1}{2\pi} \int_0^\infty dx \, G_-(x)  e^{-x^2/4} + \frac{l_B}{2\pi l}G_+(0) + O(l_B^2)
  \end{align}
while the third one is
\begin{align}
 \int_0^\infty dx \,   x G(x)^2 & =  \frac{1}{2}  \int_0^\infty dx \,  x \left[G_+^2(x) + G_-^2(x) \right]   + O(l_B^2)
  \end{align} \\
  A closer look indicates that the edge contribution to the even cumulants $\Delta \kappa_{2n}$ are $O(1)$, while it is $O(l_B)$ for the odd ones $\Delta \kappa_{2n-1}$ (with $n \geq 2$ in both cases).  

Let us finally come to the derivation of (\ref{eq:themostgeneral}), for $\textrm{Tr}\, h(V)$. Assuming that $h$ is analytic in some neighborhood of $[0,1]$ with $h(0)=0$, it can be expanded as a series $h(z)=\sum_{p\geq 1}a_p z^p$, which means it is sufficient to compute separately each $\textrm{Tr} \,V^p$. We extract this  from the asymptotic expansion of the FCS (it is also possible to get it by performing a similar analysis as we just did, but starting from (\ref{eq:importantintermediate})). We obtain
\begin{equation}
 \textrm{Tr}\, V^p=\frac{b-a}{2\pi l_B}\frac{l}{l_B}+\frac{b-a}{\pi l_B}\int_\mathbb{R} \left[\textrm{er}(x)^p-\textrm{er}(x)\right]-\frac{H_{p-1}}{\pi^2}\log\frac{l}{l_B}+O(1),
\end{equation}
for $p\geq 1$. We have used $\log(1-\omega)^2=2\sum_{p=1}^\infty H_{p-1}\omega^p/p$, where $H_n$ is the $n$-th harmonic number. Using the integral representation $H_{p-1}=\int_0^1 dz \frac{z-z^p}{z(1-z)}$, we obtain $\sum_{p\geq 1} a_p H_{p-1}=\int_0^1 dz \frac{zh(1)-h(z)}{z(1-z)}$, and (\ref{eq:themostgeneral}) follows.

\section{Second cumulants for multiple intervals in the half-plane geometry}
\label{app:C1}
In this appendix we derive formula \eqref{eq_multiple_intervals_half_plane} by computing the following asymptotic behavior of the  edge variance $\Delta \kappa$ 
\begin{align}
\label{eq_multiple_intervals_half_planeA}
\Delta \kappa & = \frac{1}{2\pi^2} \log \prod_{a=1}^p \frac{v_a -u_a}{l_B} \prod_{a<b} \frac{(u_b - v_a)(v_b - u_a)}{(u_b-u_a)(v_b-v_a)}  + n \frac{\log 2+\gamma}{4 \pi ^2} + o(1)
\end{align} 
From the discussion in the main text we have
\begin{align}
\Delta \kappa & =   \int_{\mathcal{A} \times\mathcal{A}^c} dy_1 dy_2 \,  \mathcal{K}_{l_B}(y_1-y_2), 
\end{align}
Decomposing $ \mathcal{A} \times\mathcal{A}^c = \cup_{a=1}^p \left( I_a \times I_a^c  \right)   \setminus \cup_{a \neq b} \left( I_a \times I_b \right)$ we find
 \begin{align}
\Delta \kappa & =   \sum_{a=1}^p \int_{I_a \times I_a^c} dy_1 dy_2 \,  \mathcal{K}_{l_B}(y_1-y_2) - \sum_{a\neq b} \int_{I_a \times I_b} dy_1 dy_2 \,  \mathcal{K}_{l_B}(y_1-y_2)
\end{align}
The first term is simply the sum of the entropies of the single interval $I_a$, whose asymptotic we have already calculated (up to and including $O(l_B^2)$ corrections). So we only have to compute the asymptotic of 
 \begin{align}
F_{a,b} = \int_{I_a \times I_b} dy_1 dy_2 \,  \mathcal{K}_{l_B}(y_1-y_2), \qquad a \neq b
\end{align}
However this is very easy as it is $o(1)$, and one can simply put $l_B =0$ to get the leading asymptotic.
 \begin{align}
F_{a,b}  & = \int_{I_a \times I_b} dy_1 dy_2 \,  \frac{1}{4 \pi^2 (y_1 - y_2)^2} + o(1)   = \frac{1}{4 \pi^2 } \log  \frac{(v_b - v_a)(u_b - u_a)}{(u_b- v_a)(v_b - u_a)}  + o(1) 
\end{align}
assuming without loss of generality $u_a < v_a < u_b < v_b$. Putting everything back together we get \eqref{eq_multiple_intervals_half_planeA}.

\section{Edge corner contributions}
\label{app:corner}

We comment here on a slightly generalized geometry, in which the region $A$ intersects the quantum Hall droplet with an angle $\frac{\pi}{2} + \Omega$. 

\begin{wrapfigure}{l}{0.4\textwidth}
	\centering
	\includegraphics[width=0.3\columnwidth]{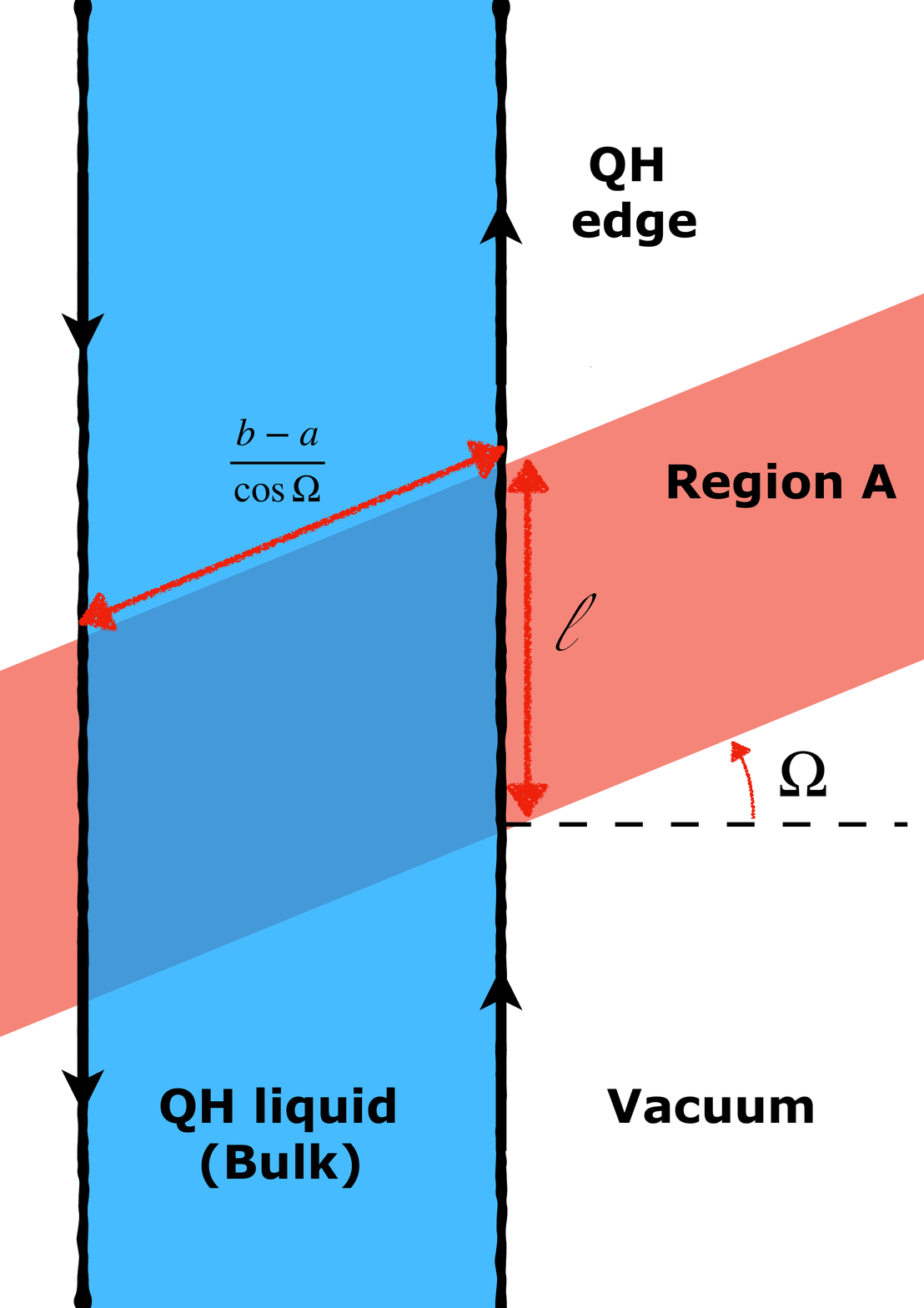}
  \caption{In red, the region $A$ used for the bipartition defining the entanglement entropy makes an angle $\pi/2+\Omega$ with the edge of the QH droplet.}
	\label{Fig:corner}
\end{wrapfigure}

Namely we consider a QH droplet on the strip $a \leq x \leq b$ as before, corresponding to the kernel 
\begin{equation}
 K(x_1,y_1;x_2,y_2)=\frac{1}{2\pi\sqrt{\pi} l^3_B} e^{- \frac{x_1^2 + x_2^2}{2l_B^2}} \int_{a}^{b} dq \, e^{\frac{q (z_1+\bar{z}_2)}{l_B^2}}   e^{-\frac{q^2}{l_B^2}} ,
\end{equation}
 but we now pick for region $A$ the following (see fig. \eqref{Fig:corner})
\begin{align*}
A = \{ (x,y),\, x \tan \Omega \leq y \leq x \tan \Omega  + l \} 
\end{align*}

We can parametrise points in $A$ as $x = t \cos \Omega$ and $y = t \sin \Omega + y$, in which case $dx \wedge dy$ becomes $\cos \Omega dt \wedge dy$. Upon integrating over $t\in \mathbb{R}$, we get 
\begin{align*}
\textrm{Tr}\, K^p &=   \left(\frac{1}{2\pi  l_B^2} \right)^p \int_{\mathcal{A}^p} d^py \,  \int_{[a,b]^p} d^pq \,  e^{i \sum_{j=1}^p \frac{y_j( q_j - q_{j-1})}{l_B^2}}    e^{- \sum_{j=1}^p  \frac{(q_j-q_{j-1})^2}{4 l_B^2 \cos^2\Omega}  } 
\end{align*}
so the 1d kernel is 
\begin{align*}
\tilde{V} (q_1,q_2)=  \frac{l }{2\pi l_B^2} e^{-\frac{(q_1-q_2)^2}{4 l_B^2 \cos^2 \Omega}}  \textrm{sinc}  \frac{(q_1-q_2) l}{2 l_B^2} 
\end{align*}
changing to $q = l_B x \cos \Omega $ we end up with a kernel
\begin{align*}
V (x_1,x_2)= f_{\Omega}(x_1-x_2), \qquad f_{\Omega}(x) =  \frac{l \cos \Omega }{2\pi l_B} e^{-\frac{x^2}{4 }}  \textrm{sinc}  \frac{x l \cos \Omega}{2 l_B} 
\end{align*}
\\

We recover the previous kernel \eqref{eq:thekernel}, up to the substitution $l \to l \cos \Omega$, $a \to a/\cos \Omega$, $b \to b/\cos \Omega$. The bulk contribution to the entropy is the expected one, namely proportional to $\frac{b-a}{\cos \Omega}$, while the edge contribution is
\begin{align}
\Delta S_n(\Omega,l) = \Delta S_n(l\cos \Omega) 
\end{align}
From  the previous analysis of the case $\Omega = 0$, we know that the edge contribution to the entropy has an asymptotic expansion of the form
\begin{align}
 \Delta S_n(l)=  \frac{1}{12}\left(1+\frac{1}{n}\right)\log \frac{l}{l_B} + O(1), \qquad (l/l_B \to \infty) 
\end{align}
Furthermore being a function of $l/l_B$, the $O(1)$ term is a pure constant, and it does not depend on $l$. This means that for an angle $\Omega$ we have   
\begin{align}
\Delta S_n(\Omega,l)  = \frac{1}{12}\left(1+\frac{1}{n}\right)\log \frac{l}{l_B} +  \frac{1}{12}\left(1+\frac{1}{n}\right)\log \cos \Omega + C + O(l_B)
\end{align}
for some constant $C$ independent of $\Omega$. The angle at which the region $A$ intersects the quantum Hall droplet contributes (at leading order) a constant term $\frac{1}{12}\left(1+\frac{1}{n}\right)\log \cos \Omega$ to the edge entropy. However it is unclear whether this result is universal or only valid in the specific geometry under consideration.

\section{Second cumulant in the disk geometry}
\label{app:C2}
In this appendix we fill the holes in the derivation of the result 
\begin{equation}\label{eq:toshow}
 \kappa_N(\theta)=\frac{\sqrt{N}}{\pi^{3/2}}+\frac{1}{2\pi^2}\log \left(\sqrt{N}\sin \frac{\theta}{2}\right)
 -\frac{1+(\pi-\theta)\cot \theta}{4\pi^2}+C+O(1/\sqrt{N}),
\end{equation}
for the fluctuations of the $N$-particle IQH state quoted in the main text.  As already explained there, the dependence on $\theta$ in this expression can be obtained by computing the second derivative wrt $\theta$, and then integrating back using the symmetry $\theta\to2\pi-\theta$ which holds for all cumulants. In this procedure an extra constant $C$ (which does not depend on $\theta$), as well as the leading behavior with $N$ is left undetermined.

We determine the leading asymptotic behavior using a different method. We first recall the fact that 
\begin{equation}
 \kappa_N(\theta)=\int_{A\times A^c} d^2 z d^2 w |V_N(z,w)|^2.
\end{equation}
Then we introduce the difference
\begin{equation}
 D_N(\theta)=\kappa_{N+1}(\theta)-\kappa_N(\theta),
\end{equation}
and look for an asymptotic expansion of $D_N$. From (\ref{eq:toshow}) we expect an expansion of the form $D_N(\theta)=a N^{-1/2}+bN+O(N^{-3/2})$ where $a=\frac{1}{2\pi^{3/2}}$ and $b=\frac{1}{4\pi^2}$ are coefficients that do not depend on $\theta$. Let us now show that this is true.  Using the alternative representation $\Gamma(N,\alpha)=e^{-\alpha}\sum_{k=0}^N \alpha^k/k!$, going to polar coordinates and computing the gaussian integrals, we obtain
\begin{equation}
 D_N(\theta)=\frac{\theta}{2\pi}\left(1-\frac{\theta}{2\pi}\right)-\frac{2}{\pi^2}\sum_{k=1}^N \frac{\sin^2 \frac{\theta k}{2}}{k^2}\frac{\Gamma(1+N-k/2)^2}{N!(N-k)!}
\end{equation}
after some algebra (an alternative method is to start from (\ref{eq:fcsqhediscrete}) instead). Now, since $\sum_{k=1}^{\infty} \frac{\sin^2 \frac{\theta k}{2}}{k^2}=\frac{\theta}{8}\left(2\pi-\theta\right)$, the above may be rewritten as
\begin{equation}
 D_N(\theta)=\frac{2}{\pi^2}\left[\sum_{k=N}^{\infty} \frac{\sin^2 \frac{\theta k}{2}}{k^2}-\sum_{k=1}^N \frac{\sin^2 \frac{\theta k}{2}}{k^2}\left(\frac{\Gamma(1+N-k/2)^2}{N!(N-k)!}-1\right)\right]
\end{equation}
The first term is easily handled by writing $\sin^2 \frac{\theta k}{2}=\frac{1}{2}-\frac{\cos \theta k}{2}$, and noticing that the second term is subleading in any asymptotic expansion, since it oscillates. We find
\begin{equation}
 \sum_{k=N}^{\infty} \frac{\sin^2 \frac{\theta k}{2}}{k^2}=\frac{1}{2N}+O(\frac{1}{N^2}).
\end{equation}
The second one,
\begin{equation}
 \tilde{D}(\theta,N)=\sum_{k=1}^N \frac{\sin^2 \frac{\theta k}{2}}{k^2}\left(\frac{\Gamma(1+N-k/2)^2}{N!(N-k)!}-1\right),
\end{equation}
is tougher. First, let us observe that it does converge to zero when $N\to \infty$. This is because the factor inside the bracket is small for $k\ll N$, while the other is otherwise. The ratio of Gamma functions gives a non trivial contribution when $k\lessapprox \sqrt{N}$. For $k$ of order $N^{1/2+\epsilon}$, it is exponentially small in $N^{\epsilon}$ for any $\epsilon>0$. Using the Stirling asymptotic expansion for the gamma function, we obtain
\begin{equation}\label{eq:someexpansion}
 \frac{\Gamma(1+N-k/2)^2}{N!(N-k)!}=e^{-k^2/(4N)}\left(1-\frac{k^3}{8N^2}+\ldots\right),
\end{equation}
where the remaining terms are a series in $N^{-2q}P_q(k)$, where $P_q$ is polynomial of degree $3q$ in $k$, $q \geq 2$. We now use the approximation
\begin{equation}
 \tilde{D}(\theta,N)=\sum_{k=1}^N \frac{\sin^2 \frac{\theta k}{2}}{k^2}\left(-1+e^{-k^2/(4N)}\left[1-\frac{k^3}{8N^2}\right]\right)+\ldots
\end{equation}
and will justify later that the unwritten terms in the above equation contribute to $O(N^{-3/2})$ and can be neglected. Similar to before we write $\sin^2 \frac{\theta k}{2}=\frac{1}{2}-\frac{\cos \theta k}{2}$, to cut the sum into two terms $\tilde{D}(\theta,N)=\tilde{D}_1(N)+\tilde{D}_2(\theta,N)$. We have
\begin{align}
 \tilde{D}_1(N)&=\frac{1}{2}\sum_{k=1}^N \frac{1}{k^2}\left(-1+e^{-k^2/(4N)}\left[1-\frac{k^3}{8N^2}\right]\right)+O(N^{-3/2})\\
 &=\frac{1}{4}\left(-1+e^{-1/(4N)}\right)+\frac{1}{2}\int_1^N \frac{dk}{k^2} \left(-1+e^{-k^2/(4N)}\left[1-\frac{k^3}{8N^2}\right]\right)+O(N^{-3/2})\\
 &=-\frac{1}{16N}+\frac{1}{4\sqrt{N}}\int_{1/2\sqrt{N}}^{\sqrt{N}/2}\frac{dq}{q^2}\left(-1+e^{-q^2}\right)-\frac{(2\sqrt{N})^2}{16N^2}\int_{1/2\sqrt{N}}^{\sqrt{N}/2} qdq e^{-q^2}+O(N^{-3/2})\\
 &=-\frac{1}{16N}+\frac{1}{4\sqrt{N}}\left(\int_0^\infty-\int_{\sqrt{N}/2}^\infty-\int_{0}^{1/2\sqrt{N}}\right) \frac{dq}{q^2}\left(-1+e^{-q^2}\right)-\frac{1}{4N}\int_0^\infty q dq e^{-q^2}+O(N^{-3/2})\\\label{eq:contri1}
 &=-\frac{\sqrt{\pi}}{4\sqrt{N}}+\frac{7}{16N}+O(N^{-3/2})
\end{align}
where we have used the Euler-Maclaurin formula in the second line. We are now also able to treat the other higher order corrections steming from  (\ref{eq:someexpansion}), which give terms of order $\frac{(\sqrt{N})^{3q-1}}{N^{2q}}=N^{-(1+q)/2}$. Doing a similar analysis at $\theta=\pi$, and using the fact that the result should not depend on $\theta$, we also obtain:
\begin{equation}\label{eq:contri2}
 \tilde{D}_2(\theta,N)=-\frac{1}{16N}+O(N^{-3/2}),
\end{equation}
A (more satisfactory) proof of (\ref{eq:contri2}), not relying on any particular value of $\theta$ is also is presented on the next page, for completeness. Summing the two contributions (\ref{eq:contri1}),(\ref{eq:contri2}) we get
\begin{equation}
 \tilde{D}(\theta,N)=-\frac{\sqrt{\pi}}{4\sqrt{N}}+\frac{3}{8N}+O(N^{-3/2}).
\end{equation}
Hence
\begin{equation}
 D_N(\theta)=\frac{1}{2\pi^{3/2}\sqrt{N}}+\frac{1}{4\pi^2 N}+O(N^{-3/2})
\end{equation}
Using this asymptotic result, it is easy to reconstruct $\kappa_N(\theta)$, using
\begin{align}
 \kappa_N(\theta)-\kappa_1(\theta)&=\sum_{k=1}^{N-1}D_N(\theta)\\
 &=\sum_{k=1}^N\left(\frac{1}{2\pi^{3/2}\sqrt{N}}+\frac{1}{4\pi^2 N}\right) +\sum_{k=1}^N O(N^{-3/2})
\end{align}
The series on the rhs is convergent and depends on $\theta$, while the other is easy to treat. In the end
\begin{equation}
  \kappa_N(\theta)=\frac{\sqrt{N}}{\pi^{3/2}}+\frac{1}{4\pi^2}\log N +B(\theta)+O(N^{-1/2}).
\end{equation}
The order one term $B(\theta)$ is left undetermined by this method, but it depends a priori on $\theta$. The dependence on $\theta$ can be fixed as explained in the main text, leading finally to (\ref{eq:toshow}).

\textbf{The remaining term.} We finally come back to the sum
\begin{equation}
 \tilde{D}_2(\theta,N)=\frac{1}{2}\sum_{k=1}^N \frac{\cos \theta k}{k^2}\left(1-e^{-k^2/(4N)}\right)
\end{equation}
and show that the leading term in the asymptotic expansion does not depend on $\theta$. We once again cut the sum in two. The first one may be treated by using the identity
\begin{equation}
 \sum_{k=1}^\infty \frac{\cos \theta k}{k^2}=\frac{\pi^2}{6}+\frac{\theta^2}{4}-\frac{\pi |\theta|}{2}
\end{equation}
for $\theta \in [-2\pi,2\pi]$, to obtain
\begin{equation}
 \sum_{k=1}^N \frac{\cos k\theta}{k^2}=\frac{\pi^2}{6}+\frac{\theta^2}{4}-\frac{\pi |\theta|}{2}-\sum_{k=N+1}^\infty \frac{\cos k\theta}{k^2}
\end{equation}
The sum on the rhs may be expressed in terms of the Lerch Zeta function. Using known asymptotic expansions of this function, or saddle point on the integral representation, we obtain
\begin{equation}
 \sum_{k=1}^N \frac{\cos k\theta}{k^2}=\frac{\pi^2}{6}+\frac{\theta^2}{4}-\frac{\pi |\theta|}{2}+O(N^{-2})
\end{equation}
The second part is given by
\begin{align}
 \sum_{k=1}^{N}\frac{\cos \theta k}{k^2}e^{-k^2/(4N)}&=\sum_{k=1}^{\infty}\frac{\cos \theta k}{k^2}e^{-k^2/(4N)}+O(N^{-\infty})\\
 &=\sqrt{\frac{N}{\pi}} \int_0^\infty e^{-N x^2}\sum_{k=1}^\infty \frac{\cos k(x-\theta)+\cos k(x+\theta)}{k^2}+O(N^{-\infty})\\
 &=\frac{1}{8N}+\frac{\pi^2}{6}+\frac{\theta^2}{4}-\frac{\pi\theta}{2}\textrm{erf}(\theta\sqrt{N})+O(N^{-\infty})\\
 &=\frac{1}{8N}+\frac{\pi^2}{6}-\frac{\pi \theta}{2}+\frac{\theta^2}{4}+O(N^{-\infty})
\end{align}
Finally, putting the two terms back together gives (\ref{eq:contri2}).
\twocolumngrid
\bibliography{biblio.bib}{}

\end{document}